\documentclass{aa}  
\usepackage{graphicx,natbib}
\usepackage{txfonts}
\newcommand{\nc}{\newcommand}

\nc{\RAJ}[4]{$\alpha(J2000) = {#1}^{\rm h}{#2}^{\rm m}{#3}\fs{#4}$}
\nc{\DecJ}[4]{$\delta(J2000) = {#1}\degr {#2}\arcmin {#3}\farcs{#4}$}

\begin{document}

   \title{Resolving the extended stellar atmospheres of asymptotic giant branch
     stars at (sub)millimetre wavelengths}

  \titlerunning{Resolving AGB stars at (sub)millimetre wavelengths}


   \author{W.~H.~T. Vlemmings
          \inst{1}\fnmsep\thanks{wouter.vlemmings@chalmers.se}
          \and
          T. Khouri\inst{1} \and H. Olofsson\inst{1}
          }

   \institute{Department of Space, Earth and Environment, Chalmers University of Technology, Onsala Space Observatory, 439 92 Onsala, Sweden
}

   \date{21-Feb-2019}

 
  \abstract
   {The initial conditions for mass loss during the asymptotic
     giant branch (AGB) phase are set in their extended atmospheres,
     where, among others, convection and pulsation driven shocks determine the
     physical conditions.
      }
   {High resolution observations of AGB
     stars at (sub)millimetre wavelengths can now directly determine
     the morphology, activity, density, and temperature close to the stellar photosphere.}
   {We used Atacama Large Millimeter/submillimeter Array (ALMA) high
     angular resolution observations to resolve the extended
     atmospheres of four of
     the nearest AGB stars: W~Hya, Mira~A, R~Dor, and R~Leo. We
     interpreted the observations using a parameterised atmosphere model.}
   {We resolve all four AGB stars and determine the brightness
     temperature structure between $1$ and $2$ stellar radii. For W~Hya and R~Dor we
     confirm the existence of hotspots with brightness temperatures
     $>3000$ to $10000$~K. All four stars show deviations from
     spherical symmetry. We find variations on a timescale
     of days to weeks, and for R~Leo we directly measure an
     outward motion of the millimetre wavelength surface with a
     velocity of at least $10.6\pm1.4$~km~s$^{-1}$. For all objects but W~Hya we find that
     the temperature-radius and size-frequency relations require the
     existence of a (likely inhomogeneous) layer of enhanced
     opacity.}
{The ALMA observations provide a unique probe of the structure of the
  extended AGB atmosphere. We find highly variable structures of
  hotspots and likely convective cells. In the future, these
  observations can be directly compared to multi-dimensional
  chromosphere and atmosphere models that determine the temperature,
  density, velocity, and ionisation structure between the stellar
  photosphere and the dust formation region. However, our results show
that for the best interpretation, both very accurate flux calibration
and near-simultaneous observations are essential.}

   \keywords{evolved stars, stars: AGB and post-AGB, stars:
     atmospheres, stars: activity, stars: imaging, Submillimetre: stars
               }

   \maketitle
%

\section{Introduction}

Asymptotic giant branch (AGB) stars, with main-sequence masses between
$1-8$~M$_\odot$, lose significant amounts of mass in a stellar wind and
are an important source of interstellar enrichment \citep{HO18}. It is
generally assumed that radiation pressure on dust, which forms at a
few stellar radii, drives the AGB winds \citep[e.g.][]{Hoefner08}. It is less clear how
the gas is transported from the stellar surface, through an extended
stellar atmopshere, to the cooler regions where dust can form.  Hence, our current understanding of the mass-loss
processes around AGB stars depends critically on
the description of convection, pulsations, and shocks in the extended
stellar atmosphere \citep[e.g.][]{Bowen88, Woitke2006,
  Freytag2008}. Additionally, the chemistry before and during the dust
formation is also strongly affected by shocks
\citep[e.g.][]{Cherchneff06}. 

Radio and (sub)millimetre observations are uniquely suited to
determining the properties of the extended AGB atmosphere. Continuum
observations at centimetre wavelengths have shown emission in excess of stellar black-body radiation. In
\citet[][hereafter RM97]{RM97}, this emission was partly resolved
  and shown to arise from
an extended region around the star where the opacity is dominated by
electron-neutral free-free emission processes. Later radio continuum
observations further resolved this region and showed that it
encompasses most of the extended AGB atmosphere out to approximately
two stellar radii \citep{RM07, Matthews15, Matthews18}.  These
observations are inconsistent with the extended chromospheres that
were produced in early AGB atmosphere models by the outward
propagation of strong shocks \citep[e.g.][]{Bowen88}. Still,
ultraviolet line and continuum emissions that appear to originate from
an AGB chromosphere have been observed around the majority of AGB
stars \citep[e.g.][]{Johnson87, Montez17}. The UV observations and
radio observations can be reconciled if the extended atmospheres of
these cool evolved stars are inhomogeneous, where the hot chromospheric
plasma lies embedded in a cooler
extended atmosphere and has a relatively low filling factor. This has previously been suggested for the
supergiant Betelgeuse \citep{Lim98}. Recent ALMA observations that
resolved the submillimetre surface of the AGB star W~Hya also support
this hypothesis \citep{Vlemmings17}.

We present further high-angular resolution ALMA observations at
millimetre and submillimetre wavelengths to probe specifically the
structure and physical properties of the inner extended
atmosphere. These observations produce direct benchmarks for the
three-dimensional atmosphere models that currently serve as input for
mass-loss prescription models \citep[e.g.][]{Freytag17}.  We present
the stars in our sample in \S~\ref{sample} and the observations and
data analysis in \S~\ref{obs}. The results of a {\it uv} analysis are given
in \S~\ref{res}. Subsequently, we present parameterised atmosphere
models used to describe our observations in \S~\ref{models}, discuss
the observed expansion of one of our sources in \S~\ref{exp}, describe
the observed hotspots and asymmetries in \S~\ref{hot}, and analyse
larger scale excess emission in \S~\ref{excess}. We end with a number
of conclusions in \S~\ref{conc}.

\begin{table*}
\caption{Observational details}             
\label{tab1}      
\centering          
\begin{tabular}{l l l l l l l l }     
\hline\hline       
Obs. date & ALMA project & Band & Ref. freq. & Beam size, P.A.$^a$ &
Rms$^{a}$ & Flux Cal.$^{b}$ & Gain cal. \\
 & & & [GHz] & [mas], [$^\circ$]  & [$\mu$Jy &  &
 \\
 & & & & & beam$^{-1}]$ & & \\
\hline                  
\multicolumn{8}{c}{Mira~A}\\
\hline 
2017-09-21 & 2016.1.00004.S & 4 & $131.28$ & $69\times57$, $46$ & $74$ & J0006-0624;~
$2.86$, $-0.55$ & J0228-0337 \\ 
2017-09-22 & 2016.1.00004.S & 6 & $213.00$ & $65\times35$, $43$ & $88$ & J0006-0624;~
$2.18$, $-0.55$ & J0228-0337 \\
2017-11-09 & 2017.1.00191.S & 7 & $330.12$ & $33\times23$, $50$ & $99$ & J0208-0047;~
$0.82$, $-0.63$ & J0217+0144 \\
\hline                  
\multicolumn{8}{c}{R~Dor}\\
\hline 
2017-09-26 & 2016.1.00004.S & 4 & $129.35$ & $74\times50$, $7$ & $75$ & J0519-4546;~
$0.93$, $-0.42$ & J0428-6438 \\ 
2017-09-25 & 2016.1.00004.S & 6 & $213.03$ & $45\times32$, $-6$ & $125$ & J0519-4546;~
$0.76$, $-0.42$ & J0428-6438 \\
2017-11-09 & 2017.1.00191.S & 7 & $330.15$ & $25\times18$, $13$ & $62$ & J0522-3627;~
$4.47$, $-0.20$ & J0516-6207 \\
\hline                  
\multicolumn{8}{c}{W~Hya}\\
\hline 
2017-10-28 & 2017.1.00075.S & 4 & $129.33$ & $48\times44$, $-65$ & $69$ & J1427-4206;~
$3.35$, $-0.65$ & J1351-2912 \\ 
2017-11-25 & 2016.A.00029.S & 7 & $330.95$ & $48\times40$, $68$ & $126$ & J1337-1257;~
$1.83$, $-0.65$ & J1351-2912 \\
\hline                  
\multicolumn{8}{c}{R~Leo}\\
\hline
2017-09-30 & 2016.1.00004.S & 4 & $129.35$ & $41\times40$, $54$ & $53$ & J0854+2006;~
$3.07$, $-0.49$ & J1002+1216 \\
2017-10-03 & 2017.1.00862.S & 6 & $215.19$ & $19\times19$, $73$ & $40$ & J0854+2006;~
$2.35$, $-0.49$ & J1002+1216 \\
2017-10-05 & 2017.1.00862.S & 4 & $139.18$ & $29\times26$, $-16$ & $42$ & J0854+2006;~
$2.91$, $-0.49$ & J1002+1216 \\
2017-10-09 & 2017.1.00862.S & 4 & $145.88$ & $38\times31$, $28$ & $29$ & J0854+2006;~
$3.02$, $-0.49$ & J1002+1216 \\
2017-10-11 & 2017.1.00075.S & 6 & $213.04$ & $33\times21$, $-46$ & $94$ & J0854+2006;~
$2.96$, $-0.49$ & J1002+1216 \\
2017-10-13 & 2017.1.00075.S & 4 & $129.34$ & $50\times35$, $31$ & $46$ & J0854+2006;~
$3.20$, $-0.49$ & J1002+1216 \\

\end{tabular}
\tablefoot{
\tablefoottext{a}{Briggs $(+0.5)$ weighted image of a single spw at the reference frequency.}
\tablefoottext{b}{Calibrator name, flux densities (Jy at the reference frequency), and
  spectral index.}
}
\end{table*}

\section{Sources}
\label{sample}

\begin{table}
\caption{Adopted parameters of the sample$^a$}             
\label{Tab:sample}      
\centering          
\begin{tabular}{l l l l }     
\hline\hline       
Star & Distance & Radius $R_*$ & Period \\
& [pc] & [mas] & [days] \\
\hline
Mira~A & 100 & 15 & 332 \\
R~Dor & 60 & 27.5 & 362/175\\
W~Hya & 98 & 20 & 388 \\
R~Leo & 71 & 13.5 & 310 \\
\end{tabular}
\tablefoot{
\tablefoottext{a}{References listed in the text}
}
\end{table}

We observed four nearby AGB stars with ALMA at high angular resolution to describe their extended atmosphere at (sub)millimetre
wavelengths. In this section, we briefly present the four stars and in particular
provide the distances used in our analysis. The observed stars pulsate
with periods of the order of one year and, over each pulsation cycle,
experience variations of orders of magnitude in visual light. The
visual-light variation is caused by the strongly time-variable
molecular opacity in this wavelength range \citep{Reid2002}.  The
opacity and, consequently, the stellar size vary strongly with
wavelength, as does the dominating opacity source (from atoms to
molecules to dust grains to free electrons).  Moreover, the stellar
size at a given wavelength is also observed to vary significantly in
time. This is a consequence of changes in density and temperature in
the inner regions that are caused by stellar pulsations and motion of
large convective cells.

\subsection{Mira~A}

Mira~A ($o$~Ceti) is the AGB star in the Mira AB system, and the
archetypal Mira variable. Its companion is most likely a white dwarf,
but a K~dwarf identification has also been suggested \citep[e.g.
  ][]{Ireland2007}. The two stars are physically separated by roughly
  90~au.  Mira~B accretes material from the circumstellar envelope of
Mira~A; there is observed interaction between the slow, dense outflow from
Mira~A and the faster, but thin outflow from Mira~B
\citep[e.g. ][]{Ramstedt2014}.  The mass-loss rate of Mira~A is
estimated to be $\sim 10^{-7}$~M$_\odot$/year
\citep{Knapp1998,Ryde2001}, but the complex morphology makes such
determinations rather uncertain.  Despite its effect on the large-scale
morphology of the circumstellar envelope, the companion is not
expected to affect the close environment of the AGB star
\citep{Mohamed2012}.  The distance to the Mira system derived
  from {\it Hipparcos} observations is $92\pm10$~pc \citep{vanLeeuwen2007}
  and the distance from the period-luminosity relation is $110 \pm
  9$~pc \citep{Haniff1995}. We adopt $100$~pc. The visual
magnitude of Mira~A varies by roughly 8$^{\rm mag}$ over the pulsation
cycle with a period of $\sim 332$~days \citep{Watson2006}.

Near-infrared observations between 1.0 and 4.0~$\mu$m (at pulsation
phase $\varphi
\sim 0.7$ and $\varphi \sim 0.0$) reveal uniform-disc diameters that
vary as a function of wavelength from $\sim 30$  to $\sim 70$~mas
\citep{Woodruff2009}. Monitoring the same wavelength range over eight
pulsation cycles shows that the uniform-disc diameter varies in time
at a given wavelength by at least 10 to 40\%
\citep{Woodruff2008}. A correction\footnote{In \citet{Vlemmings15}, an incorrect
  interpretation of the {\it uvmultifit} results meant that the
  presented axis is the minor axis instead of the major axis. As a
  result, the position angle of the major axis should be $90^\circ$
  rotated with respect to the angle presented in that paper. The
  derived brightness temperature $T_{\rm b}$ is
  lower by (axis ratio)$^2$. Hence, the results are fully consistent
  with \citet{Matthews15}.} to the average uniform-ellipse
  sizes determined by \cite{Vlemmings15} yields $(48.9 \pm
0.7) \times (43.3 \pm 0.1)$~mas at 229~GHz ($\varphi \sim 0.5$) and
of $(50.2 \pm 0.6) \times (41.8 \pm 0.4)$~mas at 94~GHz ($\varphi
\sim 0.45$). At 43 GHz, \cite{RM07} measured an uniform-ellipse
size of $54\pm5 \times 50\pm5$~mas ($\varphi = 0.05$). We used
$R_*=15$~mas, which is the smallest measurement of \citet{Woodruff2009} in the
analysis. This corresponds to $\sim1.5$~au.

\subsection{R Dor}

The period of the semi-regular variable R\,Doradus appears to vary
between two pulsation modes of 362 and 175 days on a timescale of
roughly 1000 days \citep{Bedding1998}. The non-unique period and the
modest associated visual magnitude variation ($< 2^{\rm mag}$) makes
it difficult to define the pulsation phase of R~Dor at a given time.
A distance of $\sim60$~pc is obtained from observations using {\it Hipparcos}
\citep[59\,pc,][]{Knapp2003} and using aperture-masking in the
near infrared \citep[$61\pm7$ pc;][]{Bedding1997}.
Models for observations of CO emission lines  \citep{Olofsson2002,Maercker2008,VanDeSande2018}
imply gas mass-loss rates of $\sim 10^{-7}$\,M$_\odot$\,year$^{-1}$.
Based  on oxygen isotopologue ratios, the initial stellar mass of
R~Dor is estimated to be between 1.3 and 1.6~$M_\odot$ \citep{DeBeck2018,Danilovich2017}.

Observations with SPHERE/ZIMPOL revealed the size of the stellar disc
diameter to vary strongly within the visible wavelength range, between
$\sim 58$~mas and $\sim 72$~mas \citep{Khouri16a}.  The authors also
reported a large variation in size (between $\sim 59$~mas and $\sim
72$~mas) within 48 days, which was accompanied by a striking change in
morphology.  \cite{Ireland2004} reported full width at half maximum (FWHM)
sizes between 40  and 60~mas from fitting a Gaussian source to
aperture-masked interferometry observations at visible wavelengths.
At near-infrared wavelengths, a similar technique revealed a
uniform-disc stellar radius of $\sim 27.5$~mas
\citep{Norris2012}. Hence, we adopt $R_*=27.5$~mas, or $\sim1.65$~au, in our analysis.

\subsection{W\,Hya}

The distance to W~Hydrae is found to be $\approx 100$~pc, from VLBI
astrometry of maser spots \citep[$98^{+30}_{-18}$,][]{Vlemmings2003}
and {\it Hipparcos} observations
\citep[$104^{+14}_{-11}$,][]{vanLeeuwen2007}.
Its mass-loss rate is estimated to be $\sim10^{-7}$~M$_\odot$~yr$^{-1}$, based on modelling of CO lines
\citep[e.g. ][]{Maercker2008,Khouri2014}, while its main-sequence mass
has been constrained to be between 1.0 and 1.5~M$_\odot$
\citep[][]{Khouri2014a,Danilovich2017}, based on measured oxygen
isotopic ratios.  Over a period of $\sim 388$~days, the visual
magnitude of W~Hya varies between $\sim 6^{\rm mag}$ and $9^{\rm mag}$
\citep{Zhao-Geisler2011}.  W~Hya is sometimes classified as a Mira
variable because of the relatively long period and large amplitude of its
visible magnitude variations.  However, a semi-regular classification
is more appropriate based on studies of the gas-velocity variations in
the atmosphere \citep{Lebzelter2005}.

The FWHM stellar size in the visible is found to reach values of $\sim
80$~mas at a pulsation phase of 0.44 \citep{Ireland2004}.
\cite{Ohnaka2016} and \cite{Ohnaka2017} reported observations at
maximum- and minimum-light phase, respectively, acquired in the near
infrared with AMBER/VLTI and in the visible with SPHERE/ZIMPOL. Within
one pulsation cycle, the authors measured a variation of 10\% in the
stellar size in the near infrared and a much larger variation of $\sim
50\%$ in visible wavelengths, which was accompanied by a strong change
in morphology.  As shown by \cite{Woodruff2008}, its stellar radius
also varies significantly in the near-infrared spectral range, by at
least 10 to 50\%.  The smallest reported uniform-disc diameters of
$\sim 30$~mas ($\varphi \sim 0.79$) and $\sim 40$~mas ($\varphi \sim
0.58$) were derived from observations in the near infrared by
\cite{Woodruff2009}, and the largest diameter (of $\sim 100$~mas)
obtained in the mid-infrared \citep[between $\varphi \sim 0$ and $\sim
0.5$;][]{Zhao-Geisler2011}.  ALMA images of the submillimetre continuum
emission of W\,Hya ($\varphi \sim 0.3$) revealed an asymmetric source
with size $50.9 \times 56.5$~mas and an unresolved hotspot on the
surface \citep{Vlemmings17}.  At 43~GHz, \cite{RM07} reported an
uniform-ellipse size of $(69\pm10) \times (46\pm7)$~mas (at $\varphi
\sim 0.25$) and \cite{Matthews18} found $(71 \pm 4) \times (65 \pm
3)$~mas ($\varphi \sim 0.61$).  At 22~GHz, RM97 found a uniform-disc
diameter of $80 \pm 15$~mas (average between measurements in two
consecutive pulsation cycles at $\varphi \sim 0.75$ and $\sim
0.8$). For consistency with \citep{Vlemmings17} we adopt
$R_*=20$~mas~$\approx1.96$~au.

\subsection{R~Leo}

The Mira-type variable R~Leonis experiences visible magnitude
variations of 7$^{\rm mag}$ over a period of 310 days
\citep{Watson2006}.  The distance to R~Leo is not well
  determined and published values range from $\sim70$ to
  $110$~pc. The distance from the period-luminosity relation is $110
  \pm 9$~pc \citep{Haniff1995}. The recent geometric Gaia distance is
$71^{+5}_{-4}$~pc, but the accuracy of this value is unclear since
Gaia AGB distances are highly uncertain \citep[see
e.g.][]{vL19}. Specifically for R~Leo, the Gaia result has a
goodness-of-fit value of $142$, a $\chi^2=53018$ and astrometric
excess noise of $2.7$~mas. This implies a low-quality fit and the
distance and especially the errors should thus be treated with extreme
caution. Still, we used $71$~pc in our analysis.  The mass-loss rate of
R~Leo has been determined to be $\sim 10^{-7}$~M$_\odot$/year
\citep{Danilovich2015,Ramstedt2014a}.

At visible wavelengths and $\varphi = 0.2$, the diameter was found to
vary between $48.7 \pm 2.3$ and $75.6\pm3.7$~mas depending on the
strength of TiO bands at the given wavelength \citep{Hofmann2001}.
\cite{Norris2012} reported a uniform-disc diameter of $36.6 \pm
0.6$~mas at 1.04~$\mu$m ($\varphi = 0.4$).  Observations acquired over
nine pulsation cycles showed that the uniform-disc diameter varies in
time by at least 10 to 30\% at a given wavelength in the near infrared
\citep{Woodruff2008}.  At $\varphi = 0.5$, the uniform-disc diameter
between 1.0 and 4.0~$\mu$m was observed to vary between $\sim 27$~mas
and $\sim55$~mas \citep{Woodruff2009}.
Observations by \cite{Tatebe2008} at 11.15~$\mu$m indicated a
uniform-ellipse size $(31.81 \pm 0.15) \times (30.67 \pm 0.22)$
($\varphi = 0.6$ to $0.8$).  At 43~GHz, \cite{RM07} measured the size
of R~Leo to be $(61\pm10) \times (39\pm6)$ ($\varphi \sim 0.55 $),
while \cite{Matthews18} reported $(54\pm3) \times (45\pm2)$~mas
($\varphi \sim 0.23$). We adopt the smallest measurement of
\citet{Woodruff2009}, $R_*=13.5$~mas, corresponding to $\sim0.96$~au,
as the stellar radius.

\begin{table*}
\caption{Flux density, size, temperature, and spectral index results.}             
\label{tab2}      
\centering          
\begin{tabular}{l l c c c c c c }     
\hline\hline       
Frequency  & Obs. date & $F_\nu$ & $\theta_{\rm maj}$ &
Axis ratio & P.A. &
$T_{\rm b}$ & Spectral \\
 $[{\rm GHz}]$  & & [mJy] & [mas] &  $q(=\theta_{\rm min}/\theta_{\rm
   maj})$ & [$^\circ$] & [K] & index\\
\hline                  
\multicolumn{7}{c}{Mira~A} & $2.07\pm0.16^a$ \\
\hline
$129.33$ & 2017-09-21 & $ 34.22 \pm 0.10$ & $44.59 \pm 0.60$ & $ 0.98
\pm  0.11$ & $ -40 \pm 122$ & $1845 \pm 209$ & $1.86\pm0.14$\\ 
$131.28$ & & $ 35.13 \pm 0.08$ & $45.57 \pm 0.47$ & $ 0.95 \pm  0.08$ & $ -42 \pm  47$ & $1826 \pm 157$ & \\ 
$141.42$ & $\varphi = 0.65$ & $ 40.24 \pm 0.08$ & $44.71 \pm 0.37$ & $ 0.97 \pm  0.06$ & $ -52 \pm  80$ & $1822 \pm 111$ & \\ 
$143.31$ & & $ 41.13 \pm 0.13$ & $44.62 \pm 0.65$ & $ 0.97 \pm  0.11$ & $ -13 \pm  72$ & $1835 \pm 221$ & \\ 
$213.00^b$ & 2017-09-22 & $ 83.73 \pm 0.52$ & $44.32 \pm 0.26$ & $ 0.90 \pm  0.05$ & $ -15 \pm  11$ & $1842 \pm  97$ & $^c$\\ 
$215.00^b$ & $\varphi = 0.65$ & $ 87.69 \pm 0.50$ & $42.65 \pm 0.26$ & $ 0.99 \pm  0.04$ & $   4 \pm  18$ & $1853 \pm  73$ & \\ 
$227.21^b$ & & $ 98.66 \pm 0.46$ & $43.38 \pm 0.20$ & $ 0.95 \pm  0.04$ & $ -25 \pm  17$ & $1880 \pm  76$ & \\ 
$229.13^b$ & & $ 99.61 \pm 0.54$ & $43.26 \pm 0.22$ & $ 0.95 \pm  0.04$ & $ -26 \pm  21$ & $1872 \pm  87$ & \\ 
$330.12^b$ & 2017-11-09 & $232.89 \pm 0.41$ & $42.36 \pm 0.08$ & $
0.98 \pm  0.01$ & $ -18 \pm  15$ & $2141 \pm  33$ & $1.75\pm0.14$ \\ 
$332.00^b$ & $\varphi = 0.81$ & $234.21 \pm 0.32$ & $42.55 \pm 0.07$ & $ 0.98 \pm  0.01$ & $ -18 \pm  15$ & $2102 \pm  29$ & \\ 
$342.25^b$ & & $248.50 \pm 0.28$ & $42.46 \pm 0.06$ & $ 0.98 \pm  0.01$ & $ -16 \pm  12$ & $2110 \pm  27$ & \\ 
$344.12^b$ & & $249.18 \pm 0.31$ & $42.39 \pm 0.06$ & $ 0.99 \pm
0.01$ & $ -20 \pm  15$ & $2093 \pm  27$ & \\ 
\hline                  
\multicolumn{7}{c}{R~Dor} & $2.06\pm0.16^a$ \\
\hline
$129.35$ & 2017-09-26 & $ 79.16 \pm 0.11$ & $65.29 \pm 0.48$ & $ 0.98
\pm  0.04$ & $   4 \pm  43$ & $1999 \pm  76$ & $1.99\pm0.29$ \\ 
$131.29$ & $\varphi = 0.46$ & $ 81.37 \pm 0.09$ & $64.89 \pm 0.41$ & $ 0.99 \pm  0.02$ & $  17 \pm  68$ & $1999 \pm  53$ & \\ 
$141.43$ & & $ 93.15 \pm 0.09$ & $64.78 \pm 0.39$ & $ 0.98 \pm  0.02$ & $   2 \pm  28$ & $1995 \pm  50$ & \\ 
$143.32$ & & $ 95.39 \pm 0.10$ & $64.54 \pm 0.39$ & $ 0.98 \pm  0.02$ & $  10 \pm  38$ & $1998 \pm  49$ & \\ 
$213.03$ & 2017-09-25 & $211.38 \pm 0.12$ & $62.38 \pm 0.16$ & $ 0.99 \pm  0.01$ & $  -1 \pm  28$ & $2131 \pm  33$ & $^c$\\ 
$215.03$ & $\varphi = 0.46$ & $215.59 \pm 0.13$ & $62.53 \pm 0.17$ & $ 0.98 \pm  0.01$ & $  -0 \pm  20$ & $2134 \pm  34$ & \\ 
$227.24$ & & $241.21 \pm 0.12$ & $62.16 \pm 0.15$ & $ 0.99 \pm  0.01$ & $   1 \pm  20$ & $2156 \pm  30$ & \\ 
$229.16$ & & $245.36 \pm 0.12$ & $62.09 \pm 0.15$ & $ 0.99 \pm  0.01$ & $  10 \pm  24$ & $2158 \pm  29$ & \\ 
$330.15$ & 2017-11-09 & $531.57 \pm 0.13$ & $62.29 \pm 0.07$ & $ 0.94 \pm  0.01$ & $  44 \pm   2$ & $2355 \pm  25$ & $2.05\pm0.05$\\ 
$332.03$ & $\varphi = 0.57$ & $539.89 \pm 0.13$ & $62.22 \pm 0.07$ & $ 0.94 \pm  0.01$ & $  43 \pm   2$ & $2370 \pm  25$ & \\ 
$342.30$ & & $573.25 \pm 0.13$ & $62.16 \pm 0.07$ & $ 0.94 \pm  0.01$ & $  43 \pm   1$ & $2375 \pm  25$ & \\ 
$344.17$ & & $579.94 \pm 0.13$ & $62.06 \pm 0.07$ & $ 0.94 \pm  0.01$ & $  44 \pm   2$ & $2376 \pm  25$ & \\ 
\hline                  
\multicolumn{7}{c}{W~Hya} & $1.87\pm0.15^a$ \\ 
\hline
$129.33$ & 2017-10-28 & $ 67.74 \pm 0.09$ & $53.45 \pm 0.32$ & $ 1.00 \pm  0.03$ & $ - $ & $2483 \pm  73$ & $1.83\pm0.01$\\ 
$131.29$ & $\varphi = 0.02$ & $ 69.60 \pm 0.06$ & $53.07 \pm 0.30$ & $ 1.00 \pm  0.02$ & $ - $ & $2488 \pm  58$ & \\ 
$141.42$ & & $ 79.71 \pm 0.06$ & $52.81 \pm 0.28$ & $ 1.00 \pm  0.02$ & $ -$ & $2511 \pm  51$ & \\ 
$143.32$ & & $ 81.72 \pm 0.07$ & $52.64 \pm 0.29$ & $ 1.00 \pm  0.02$ & $ - $ & $2524 \pm  55$ & \\ 
$330.95$ & 2017-11-25 & $391.37 \pm 0.13$ & $49.38 \pm 0.06$ & $ 0.97
\pm  0.01$ & $  90 \pm   5$ & $2668 \pm  32$ & $1.87\pm0.01$ \\ 
$332.84$ & $\varphi = 0.10$ & $392.98 \pm 0.15$ & $49.19 \pm 0.07$ & $ 0.96 \pm  0.01$ & $  90 \pm   4$ & $2694 \pm  34$ & \\ 
$342.96$ & & $421.25 \pm 0.12$ & $49.59 \pm 0.06$ & $ 0.95 \pm  0.01$ & $  90 \pm   3$ & $2691 \pm  31$ & \\ 
$344.96$ & & $421.32 \pm 0.18$ & $49.93 \pm 0.07$ & $ 0.94 \pm  0.01$
& $ -68 \pm   4$ & $2667 \pm  33$ & \\ 
\hline
\multicolumn{7}{c}{R~Leo} & $1.92\pm0.28^a$\\
\hline
$129.35$ & 2017-09-30 & $ 38.96 \pm 0.09$ & $43.20 \pm 0.33$ & $ 1.00
\pm  0.05$ & $ - $ & $2204 \pm 111$ & $1.68\pm0.03$\\ 
$131.29$ & $\varphi = 0.41$ & $ 39.73 \pm 0.07$ & $45.18 \pm 0.31$ & $ 0.91 \pm  0.04$ & $ -16 \pm   9$ & $2186 \pm 101$ & \\ 
$141.42$ & & $ 45.13 \pm 0.07$ & $43.68 \pm 0.28$ & $ 0.96 \pm  0.04$ & $  78 \pm  58$ & $2172 \pm  88$ & \\ 
$143.32$ & & $ 46.16 \pm 0.08$ & $42.23 \pm 0.27$ & $ 1.00 \pm  0.03$ & $ - $ & $2220 \pm  72$ & \\ 
$215.19^b$ & 2017-10-03 & $ 90.97 \pm 0.17$ & $43.75 \pm 0.11$ & $ 0.93 \pm  0.01$ & $  -8 \pm   4$ & $1940 \pm  34$ & $2.13\pm0.20$\\ 
$217.99$ & $\varphi = 0.42$ & $ 95.92 \pm 0.07$ & $41.80 \pm 0.10$ & $ 1.00 \pm  0.01$
& $ - $ & $2038 \pm  27$ & \\ 
$230.23$ & & $106.24 \pm 0.07$ & $42.05 \pm 0.10$ & $ 0.99 \pm  0.01$ & $ -87 \pm   4$ & $2020 \pm  31$ & \\ 
$232.53$ & & $107.83 \pm 0.07$ & $41.88 \pm 0.10$ & $ 0.98 \pm  0.01$ & $ -87 \pm  14$ & $2048 \pm  29$ & \\ 
$139.18^b$ & 2017-10-05 & $ 40.97 \pm 0.04$ & $43.76 \pm 0.24$ & $ 0.99 \pm  0.02$ & $  20 \pm   6$ & $1967 \pm  41$ & $1.40\pm0.04$\\ 
$141.07^b$ & $\varphi = 0.43$ & $ 41.65 \pm 0.03$ & $45.76 \pm 0.24$ & $ 0.92 \pm  0.01$ & $ -10 \pm   4$ & $1912 \pm  37$ & \\ 
$151.23^b$ & & $ 45.80 \pm 0.08$ & $45.07 \pm 0.24$ & $ 0.92 \pm  0.01$ & $  -9 \pm   3$ & $1883 \pm  35$ & \\ 
$153.16^b$ & & $ 46.88 \pm 0.11$ & $45.03 \pm 0.25$ & $ 0.92 \pm  0.02$ & $  -9 \pm   4$ & $1880 \pm  40$ & \\ 
$145.88$ & 2017-10-09 & $ 43.03 \pm 0.03$ & $46.01 \pm 0.24$ & $ 0.93 \pm  0.01$ & $  -3 \pm   4$ & $1819 \pm  34$ & $1.80\pm0.03$\\ 
$147.80$ & $\varphi = 0.44$ & $ 43.98 \pm 0.03$ & $45.90 \pm 0.24$ & $ 0.93 \pm  0.01$ & $  -3 \pm   4$ & $1813 \pm  31$ & \\ 
$157.98$ & & $ 49.53 \pm 0.03$ & $44.47 \pm 0.23$ & $ 0.98 \pm  0.01$ & $ -27 \pm   6$ & $1803 \pm  28$ & \\ 
$159.86$ & & $ 50.80 \pm 0.04$ & $45.64 \pm 0.24$ & $ 0.93 \pm  0.01$ & $  -0 \pm   4$ & $1798 \pm  31$ & \\ 
$213.04$ & 2017-10-11 & $ 96.56 \pm 0.16$ & $45.40 \pm 0.16$ & $ 0.92
\pm  0.02$ & $   6 \pm   7$ & $1977 \pm  54$ & $1.83\pm0.09$ \\ 
$215.04$ & $\varphi = 0.45$ & $ 98.68 \pm 0.18$ & $44.96 \pm 0.18$ & $ 0.93 \pm  0.03$ & $   7 \pm   9$ & $2009 \pm  61$ & \\ 
$227.25$ & & $109.45 \pm 0.16$ & $45.02 \pm 0.15$ & $ 0.92 \pm  0.02$ & $   4 \pm   7$ & $1995 \pm  49$ & \\ 
$229.16$ & & $110.00 \pm 0.17$ & $44.92 \pm 0.15$ & $ 0.92 \pm  0.02$ & $   6 \pm   7$ & $1989 \pm  50$ & \\ 
$129.34$ & 2017-10-13 & $ 36.59 \pm 0.09$ & $47.03 \pm 0.42$ & $ 0.92 \pm  0.05$ & $   5 \pm  19$ & $1886 \pm 114$ & $1.79\pm0.02$\\ 
$131.30$ & $\varphi = 0.46$ & $ 37.59 \pm 0.07$ & $46.12 \pm 0.37$ & $ 0.94 \pm  0.05$ & $   0 \pm  19$ & $1918 \pm  99$ & \\ 
$141.43$ & & $ 42.88 \pm 0.08$ & $45.03 \pm 0.32$ & $ 0.98 \pm  0.03$ & $ -17 \pm  21$ & $1900 \pm  64$ & \\ 
$143.33$ & & $ 44.04 \pm 0.08$ & $46.55 \pm 0.34$ & $ 0.92 \pm  0.04$ & $   0 \pm  13$ & $1886 \pm  85$ & \\ 
\end{tabular}
\tablefoot{
\tablefoottext{a}{Spectral index fit combining all epochs/bands.}  
\tablefoottext{b}{Fit restricted to baseline lengths $>2$~M$\lambda$ (Band 4) and $>3$~M$\lambda$ (Bands 6 and 7).}
\tablefoottext{c}{Combined spectral index fit with previous epoch.}}
\end{table*}

\section{Observations and data reduction}
\label{obs}

The observations presented in this paper were taken between 21
September and 25 November 2017 in some of the longest baseline ALMA
configurations with maximum baselines between 10 and 16.2~km. The
data were obtained as part of several different projects. Initially
we analysed data from the projects; 2016.1.00004.S, 2016.A.00029.S,
2017.1.00075.S (PI: Vlemmings), and 2017.1.00191.S (PI: Khouri). When
we noted an apparent increase in stellar size for R~Leo in our three
observing epochs (see \S~\ref{exp}), we retrieved archive data of
R~Leo taken between our observations as part of project 2017.1.00862.S
(PI: Kaminski).  All observations were taken in spectral line mode,
with four spectral windows (spws) with a bandwidth of $1.95$~GHz
each. The initial calibration (of the bandpass, amplitude, absolute
flux-density scale, and phase) for several of the data sets was done
using the Common Astronomy Software Applications (CASA) package
pipeline \citep{McMullin07}, while others were manually calibrated by observatory staff
using different version of CASA. Observation details, including the
flux and gain calibrators used, are given in Table.~\ref{tab1}.

We inspected the results of the initial calibrations, in particular
the absolute flux calibration and flagging. We identified a few cases
in which extra flagging was required.  Additionally,
because of the multi-epoch coverage of R~Leo in Band 4, we were able
to determine systematic flux density differences between calibrators for the
flux and gain that appeared to affect the measurements of the
star. Hence, the flux densities in these observations were recalibrated using
a renormalised flux density for the gain calibrator (J1002+1216) of $136$~mJy
at $135$~GHz. This reduced the flux densities by $\sim8\%$ in the epochs taken
on 30 September and 9 October. Based on the remaining scatter in the
R~Leo observations, we estimated a remaining absolute flux calibration
uncertainty in Band 4 of $\sim5\%$ for R~Leo. For the other sources
only a single epoch was available at each observing band and we assume
a $10\%$ uncertainty for all bands.

After the initial calibration, we performed self-calibration. For the
data sets taken as part of projects 2016.1.00004.S and 2017.1.00075.S,
one of the spws contained a strong SiO maser that was used for the
first two steps of phase self-calibration. These solutions were then
applied to all other spws. We subsequently carefully removed obvious
spectral lines and performed spectral (20 channels) and time smoothing
(10~sec) to reduce the data size. We then performed one or two,
depending on signal-to-noise (S/N) improvements, further
self-calibration steps for phase solutions on the continuum. These
solutions were also applied to all the data. No amplitude
self-calibration was performed and for the data sets without SiO
masers, only continuum self-calibration was done. The self-calibration
generally improved the S/N from $\sim100$ to $\sim1000$ or better.  In
Table~\ref{tab1} we give the rms noise and beam size for the lowest
frequency spw in the data using Briggs robust weighting
\citep{Briggs95}. The maximum recoverable scale depends on the
  shortest baselines in the antenna configuration and ranges from
$\sim0.4\arcsec$ at the longest wavelengths to $\sim0.2\arcsec$ at
the shortest wavelengths. Results for a selection of molecular lines
that were included in the observations have previously been presented
in \citet{Vlemmings18} and \citet{Khouri19}.

\section{Results}
\label{res}


\begin{figure*}[htb]
    \begin{minipage}[t]{0.5\textwidth}
       \includegraphics[width=\textwidth]{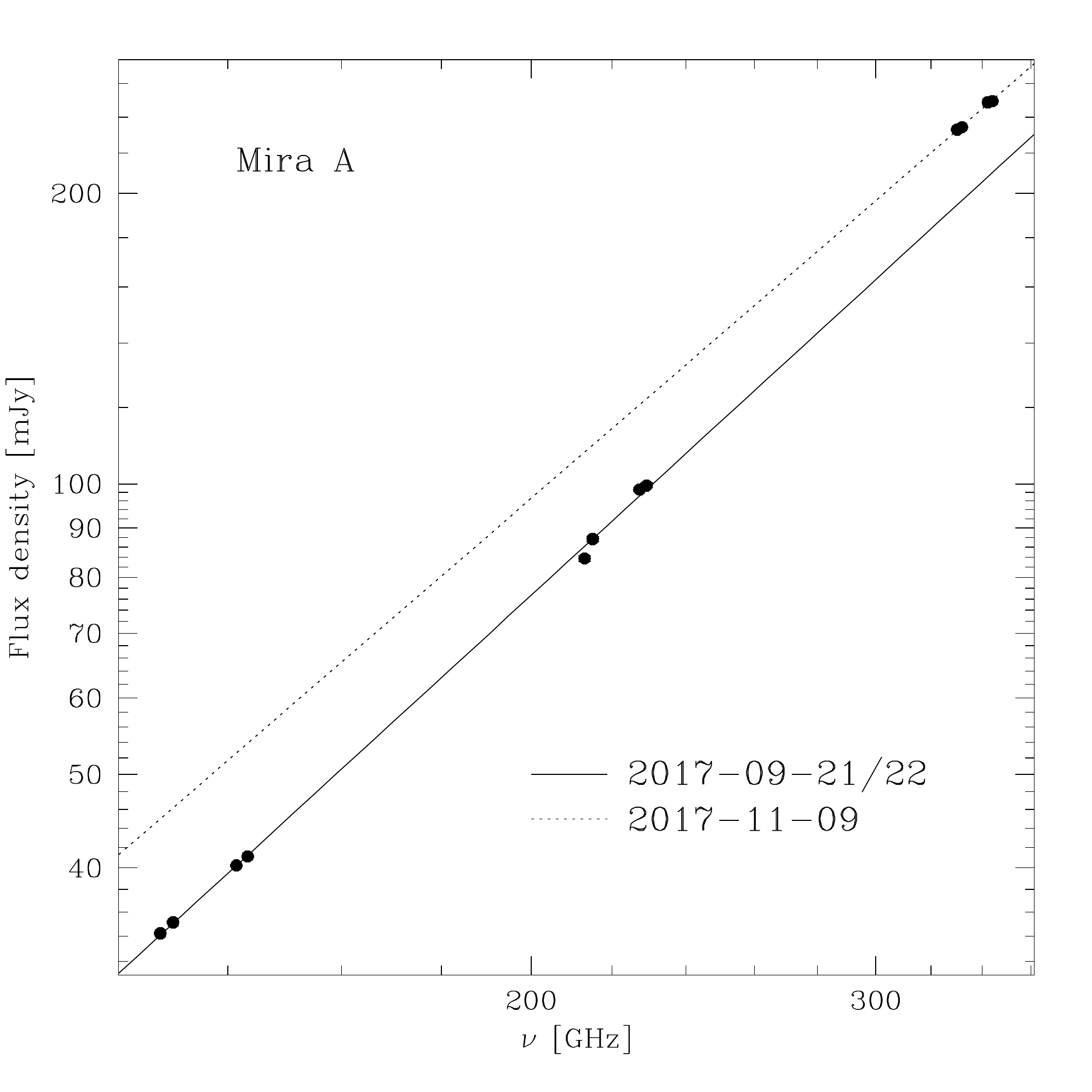}
  \end{minipage}
  \begin{minipage}[t]{0.5\textwidth}
       \includegraphics[width=\textwidth]{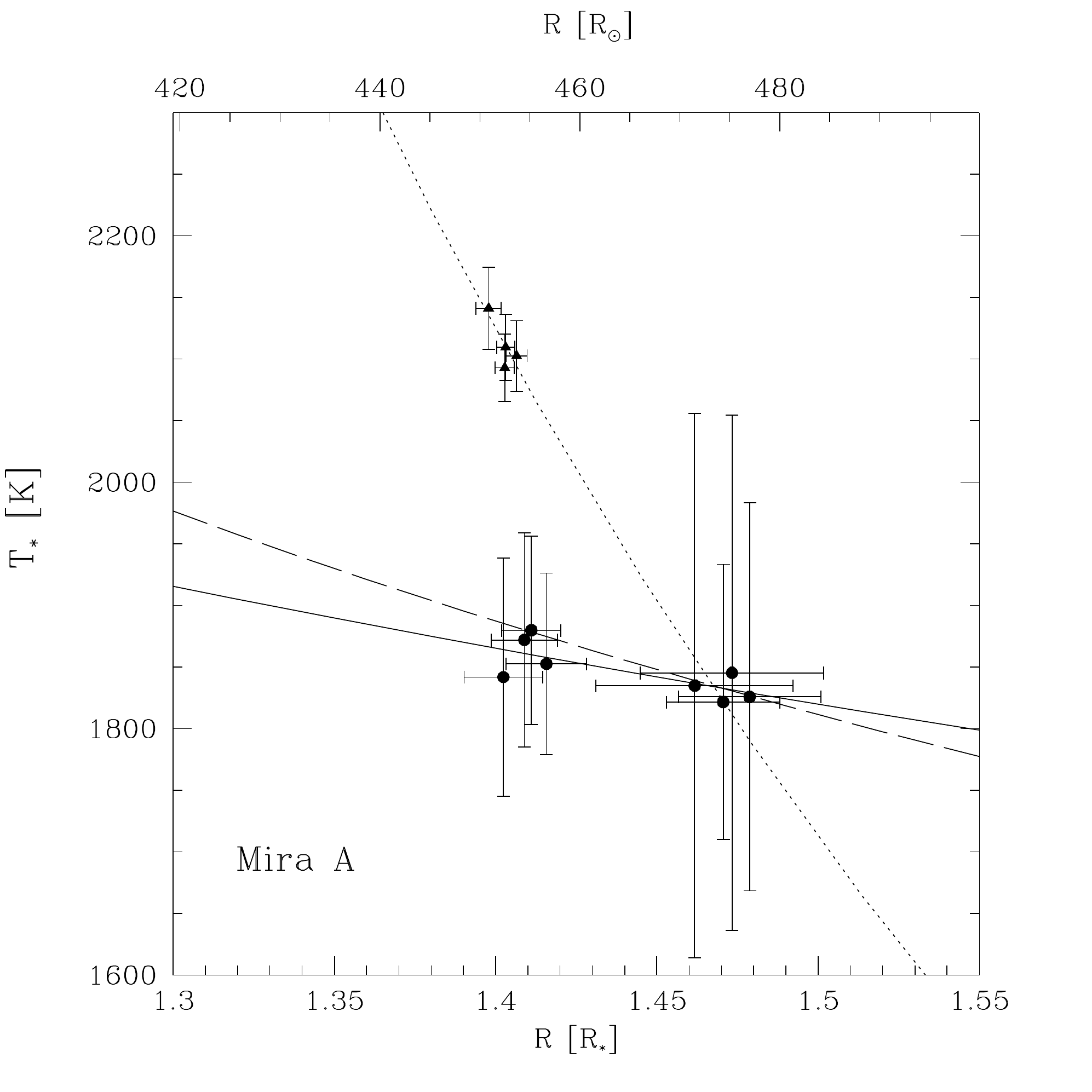}
  \end{minipage}  
\hfill
   \begin{minipage}[t]{0.5\textwidth}
       \includegraphics[width=\textwidth]{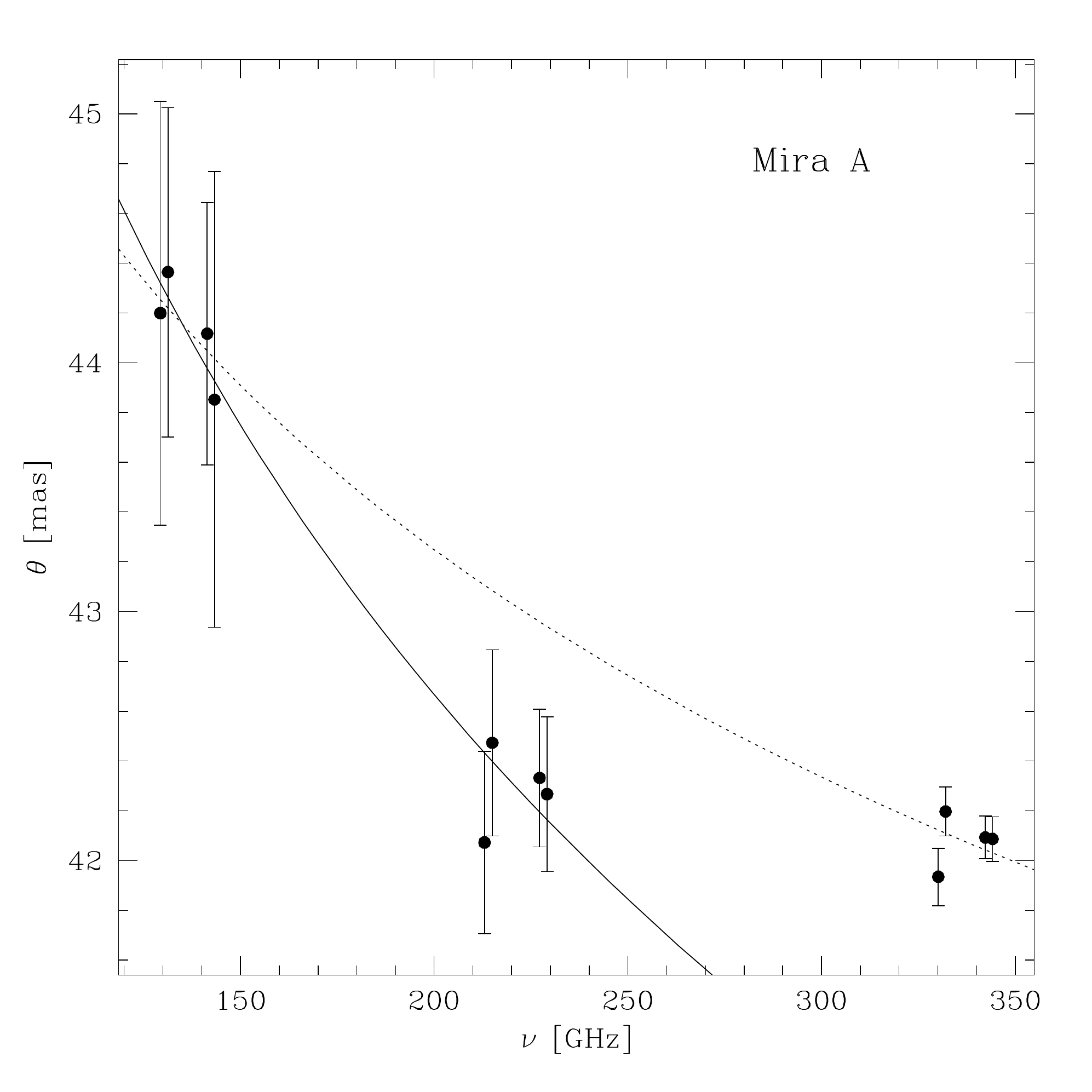}
  \end{minipage}
  \begin{minipage}[t]{0.5\textwidth}
       \includegraphics[width=\textwidth]{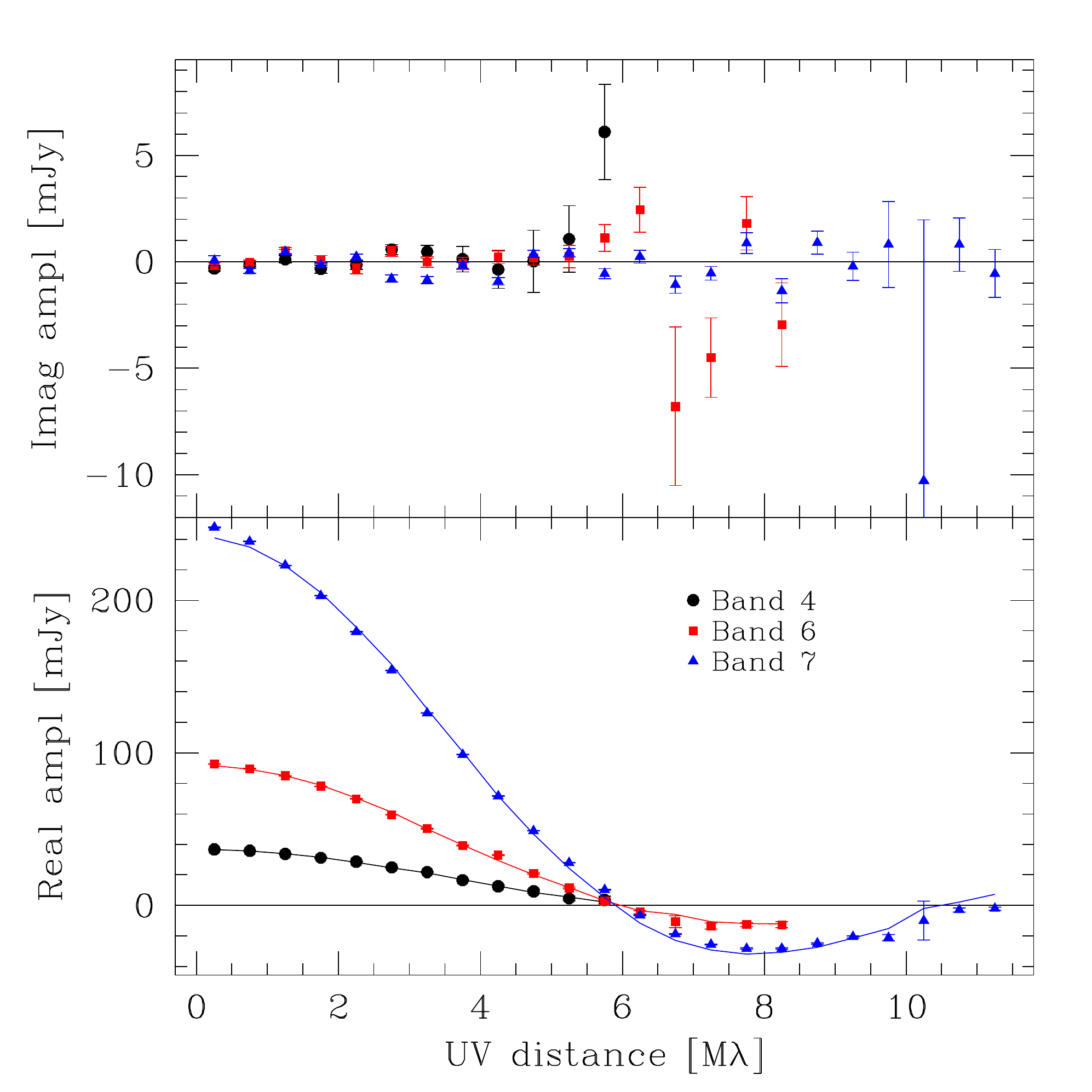}
  \end{minipage}  
   \caption{{\it Top left:} Flux density measurements determined by
      {\it uv} fitting a uniform elliptical stellar disc to Mira A. Only the fitted
      flux density uncertainties are included. The absolute flux density uncertainty
      between the different observing bands are discussed in the
      text. The symbol size
      is generally larger than the fitted flux density uncertainties. The solid and
      dotted lines represent a spectral index fit for two observing
      epochs. {\it Top right:} Derived temperature vs. radius profiles
      for Mira A. The solid and dotted lines correspond to the epochs
      in the top left figure. For illustration, the long-dashed line
      corresponds to a grey atmosphere temperature profile (RM97) with a temperature at $R_*$ of
      $T=2550$~K. {\it Bottom left:} The size vs. frequency relation at
      both observing epochs. The size $\theta$ denotes the average of
      the major and minor axis. {\it Bottom right:} The radially
      averaged real and imaginary visibilities and the best-fit
      elliptical stellar disc model of Mira A for the three observed bands.
    }
  \label{figMira}
\end{figure*}

\begin{figure*}[htb]
    \begin{minipage}[t]{0.5\textwidth}
       \includegraphics[width=\textwidth]{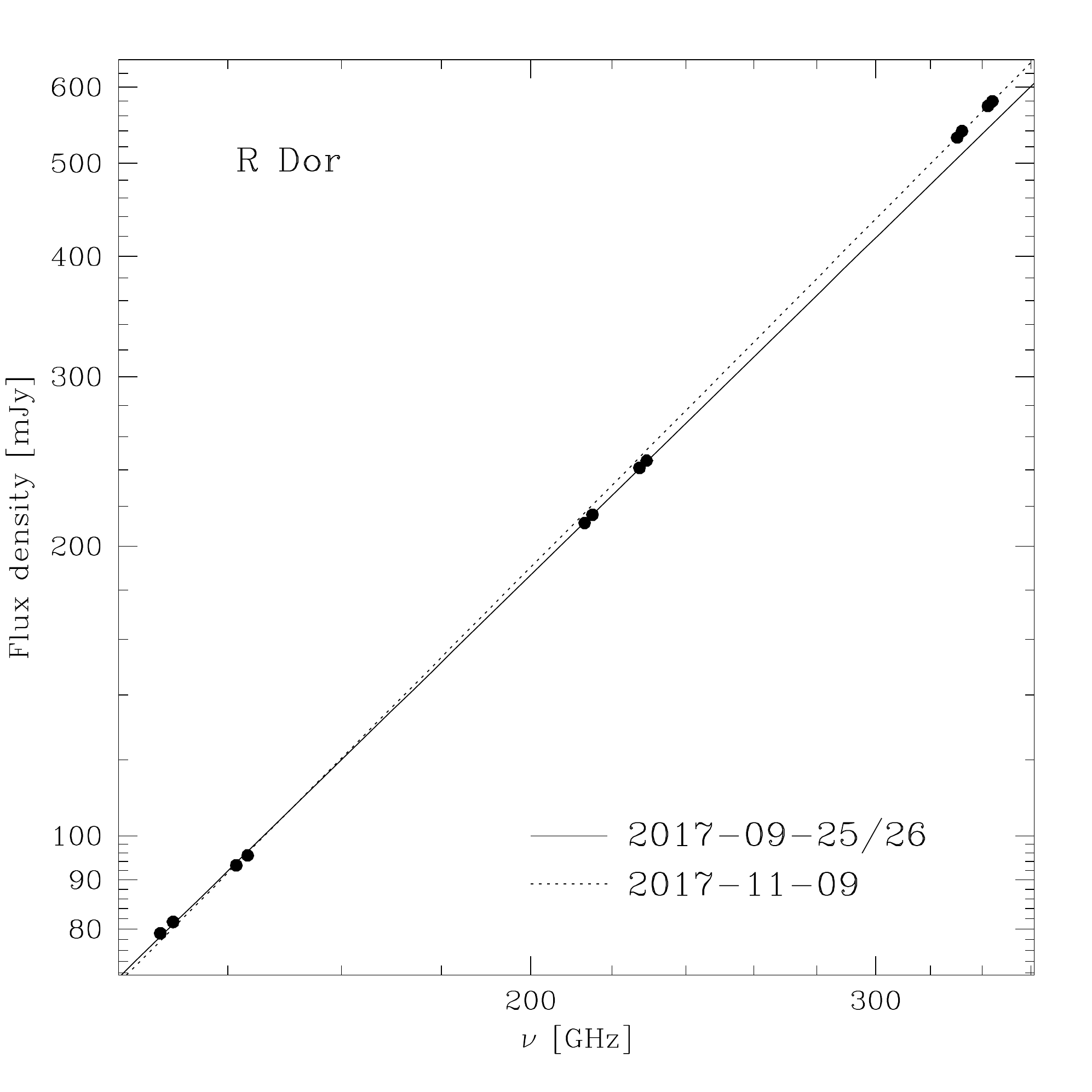}
  \end{minipage}
  \begin{minipage}[t]{0.5\textwidth}
       \includegraphics[width=\textwidth]{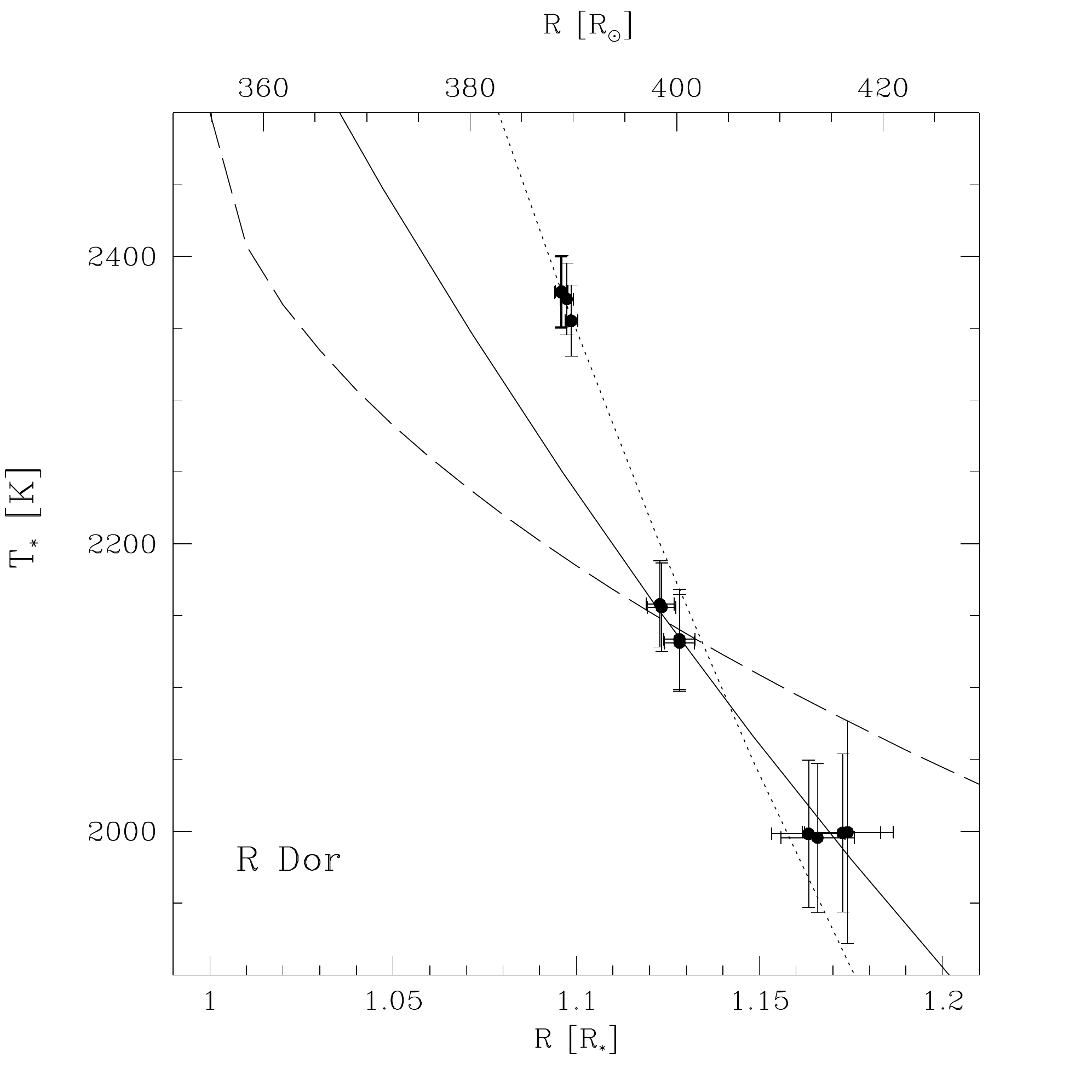}
  \end{minipage}  
\hfill
   \begin{minipage}[t]{0.5\textwidth}
       \includegraphics[width=\textwidth]{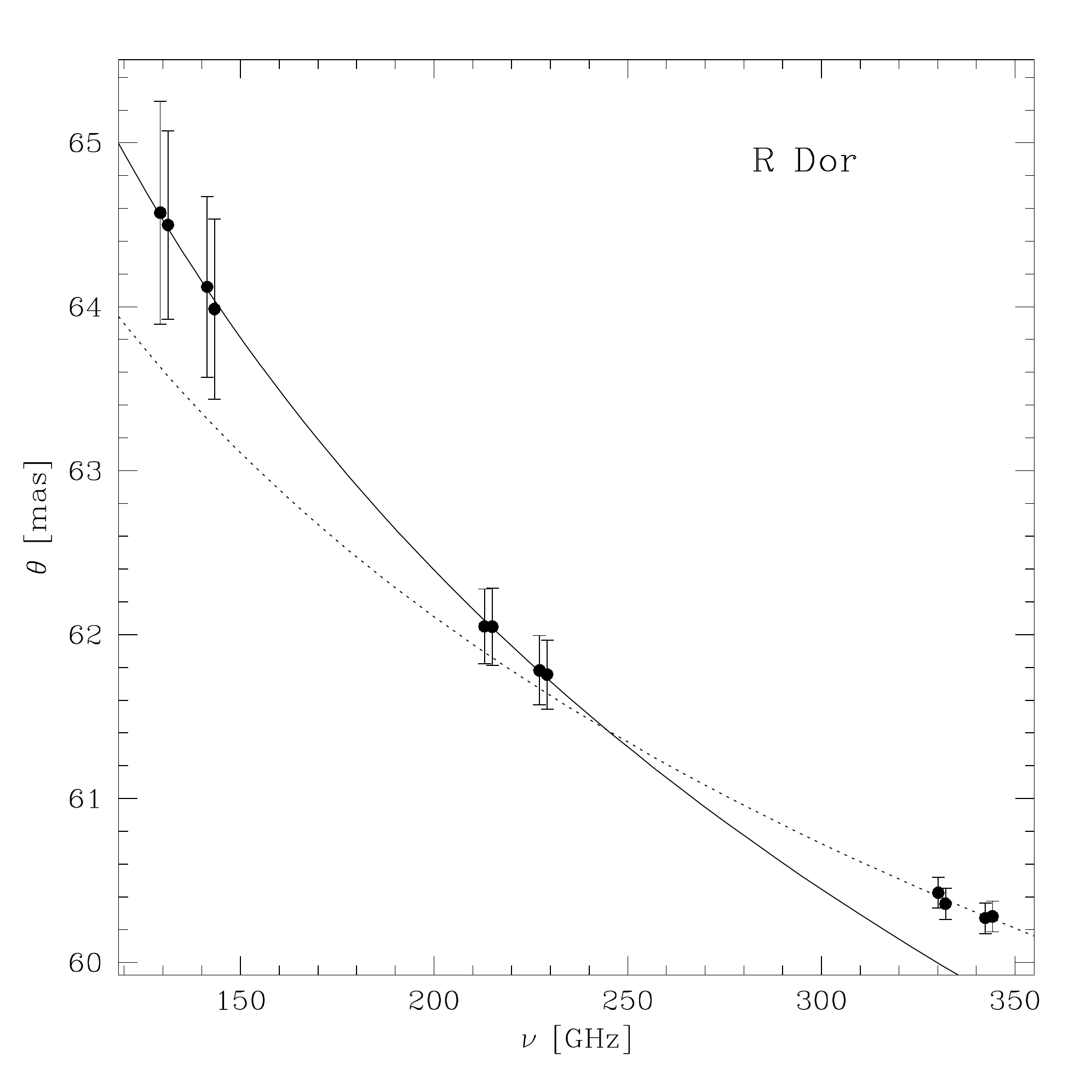}
  \end{minipage}
  \begin{minipage}[t]{0.5\textwidth}
       \includegraphics[width=\textwidth]{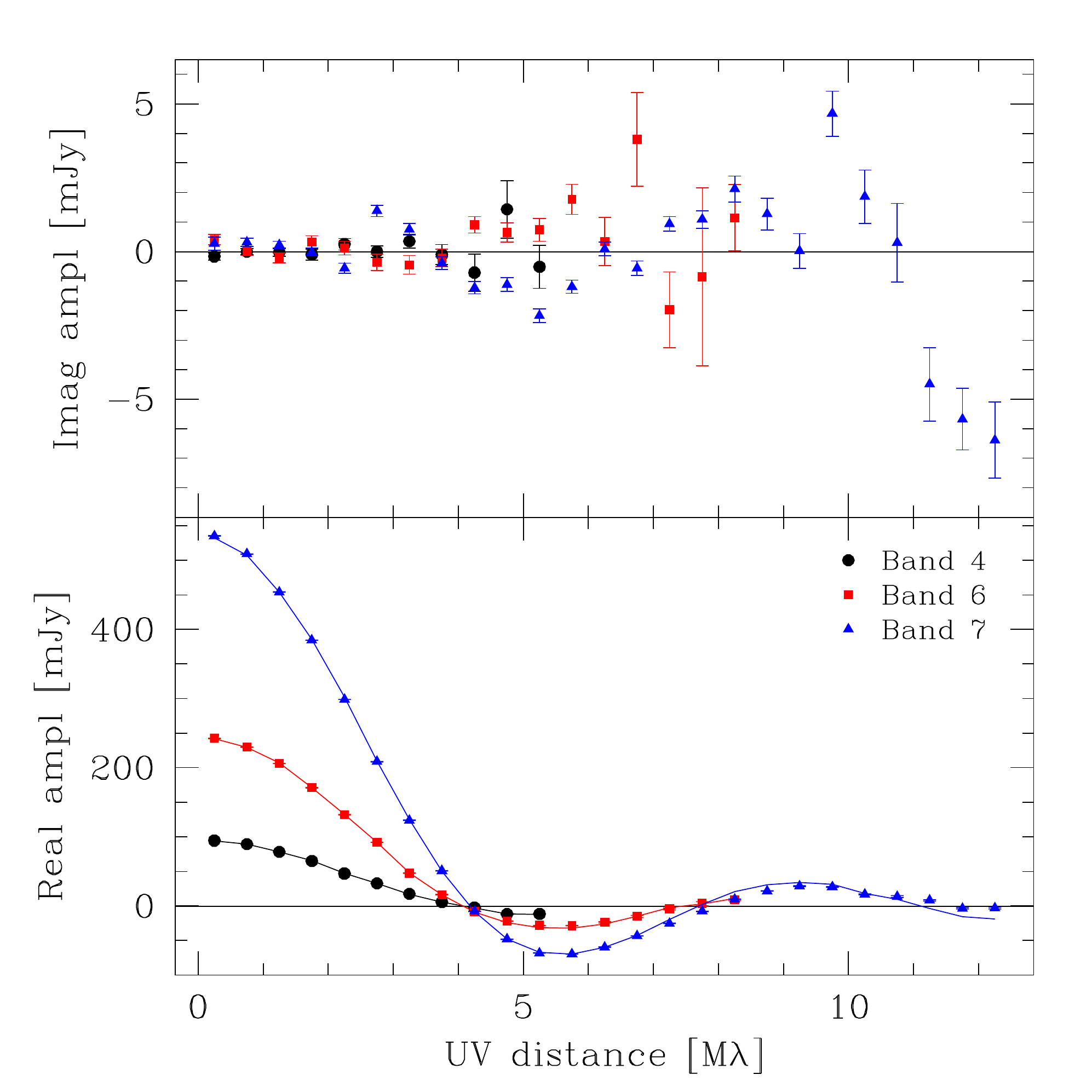}
  \end{minipage}  
   \caption{As Fig.~\ref{figMira} for R~Dor. In the top right panel,
    the long-dashed line indicates the grey atmosphere temperature profile with a temperature at $R_*$ of
      $T=2500$~K.}
  \label{figRDor}
\end{figure*}

\begin{figure*}[htb]
    \begin{minipage}[t]{0.5\textwidth}
       \includegraphics[width=\textwidth]{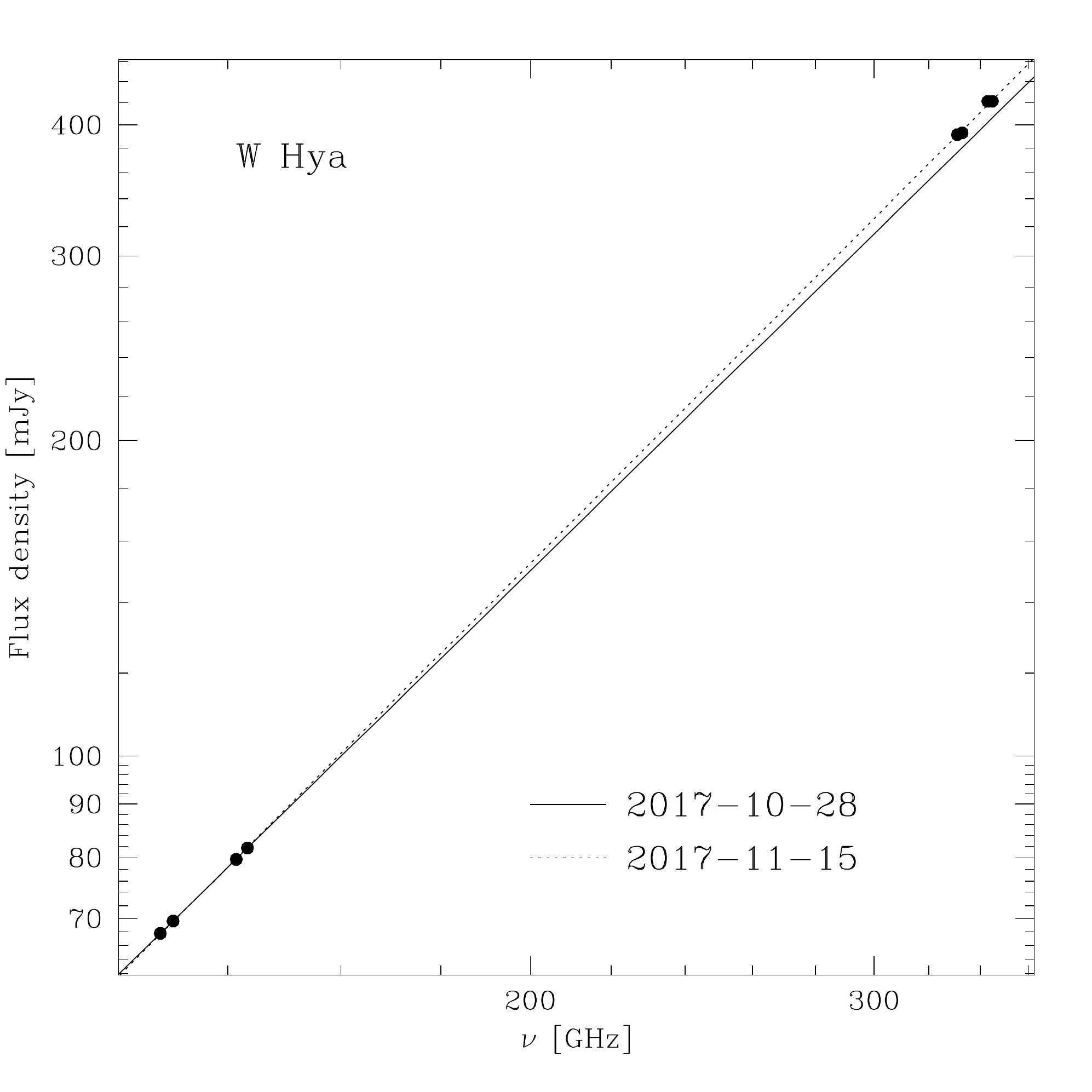}
  \end{minipage}
  \begin{minipage}[t]{0.5\textwidth}
       \includegraphics[width=\textwidth]{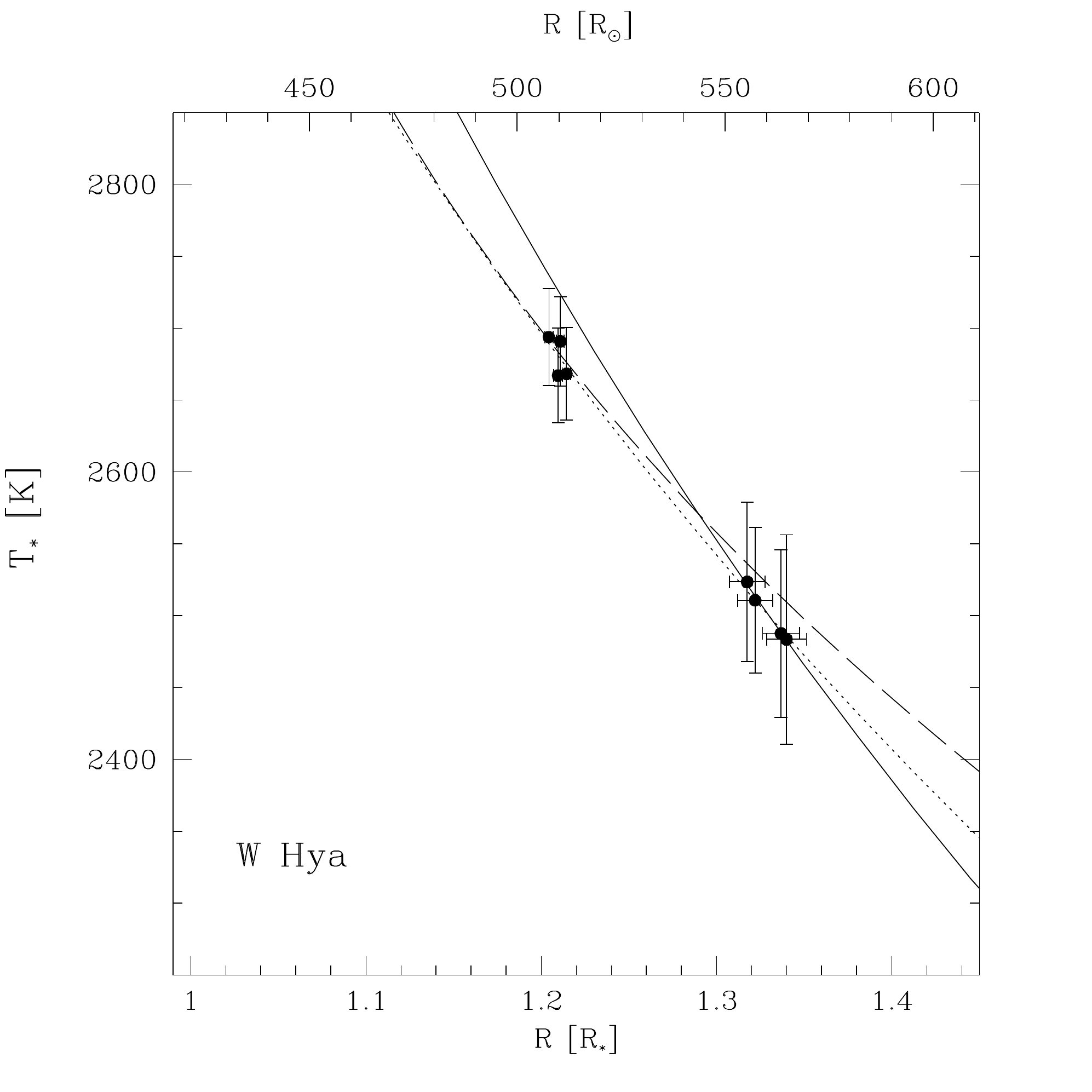}
  \end{minipage}  
\hfill
   \begin{minipage}[t]{0.5\textwidth}
       \includegraphics[width=\textwidth]{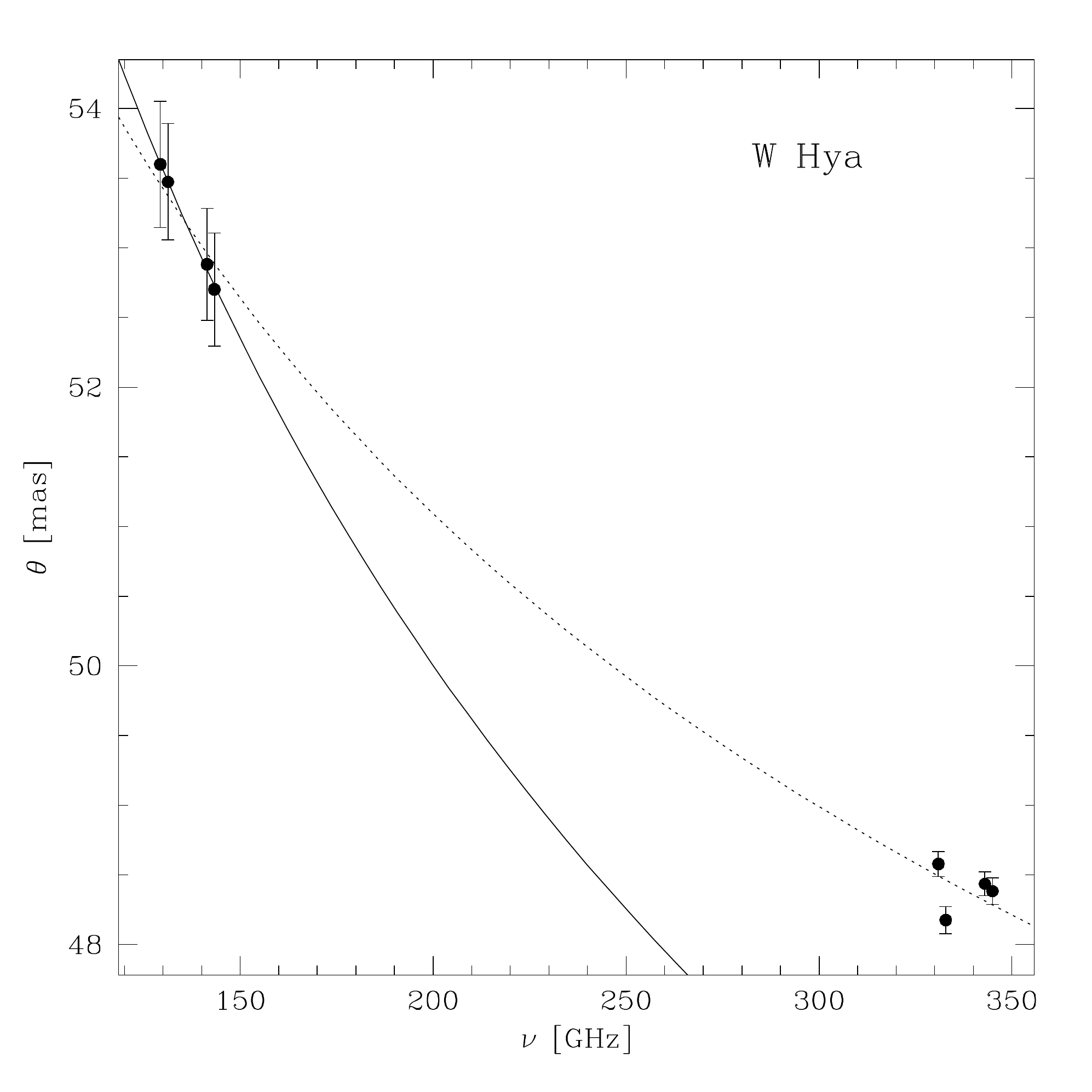}
  \end{minipage}
  \begin{minipage}[t]{0.5\textwidth}
       \includegraphics[width=\textwidth]{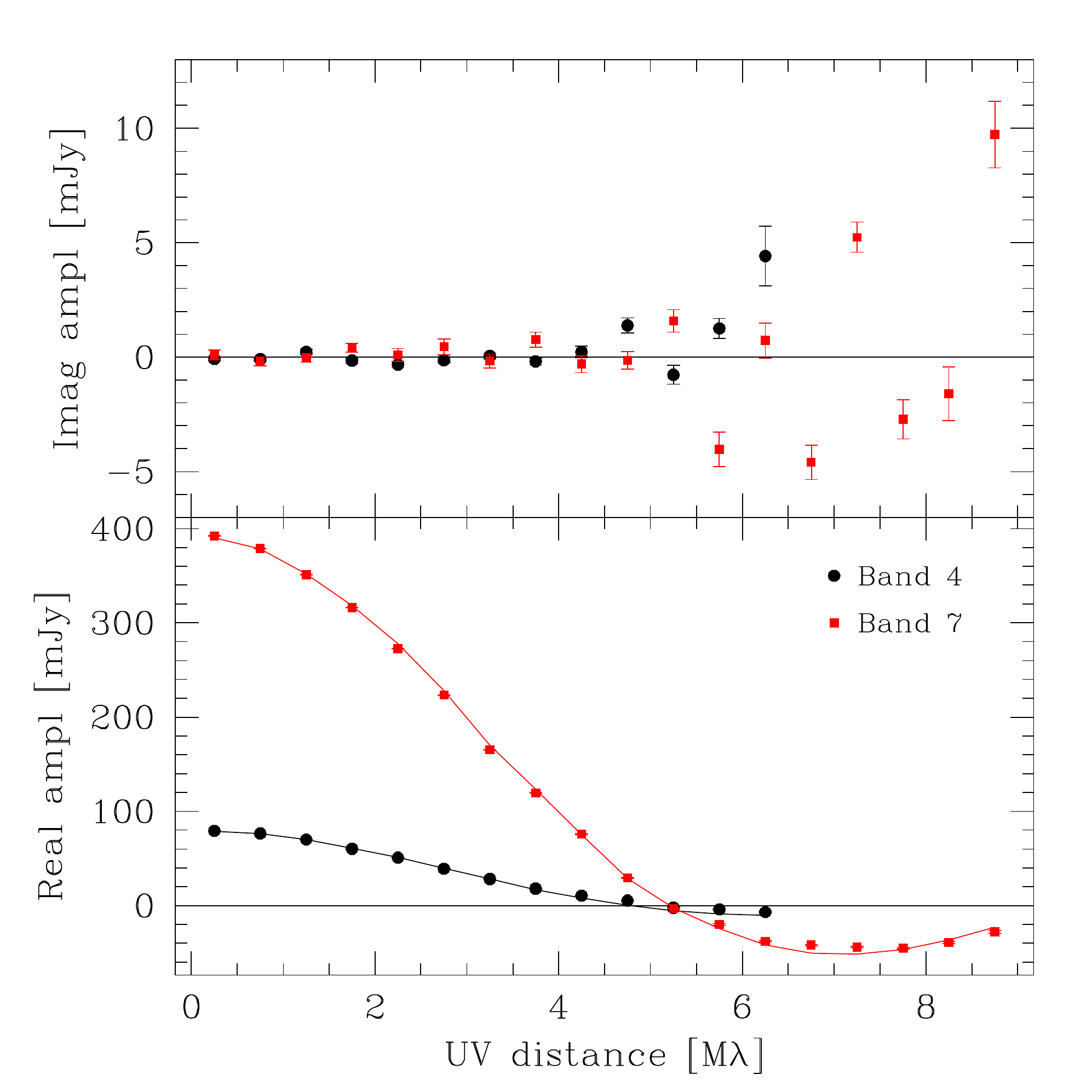}
  \end{minipage}  
   \caption{As Fig.~\ref{figMira} for W~Hya. In the top right panel,
    the long-dashed line indicates the grey atmosphere temperature profile with a temperature at $R_*$ of
      $T=3300$~K.}
  \label{figWHya}
\end{figure*}

\begin{figure*}[htb]
   \begin{minipage}[t]{0.5\textwidth}
      \includegraphics[width=\textwidth]{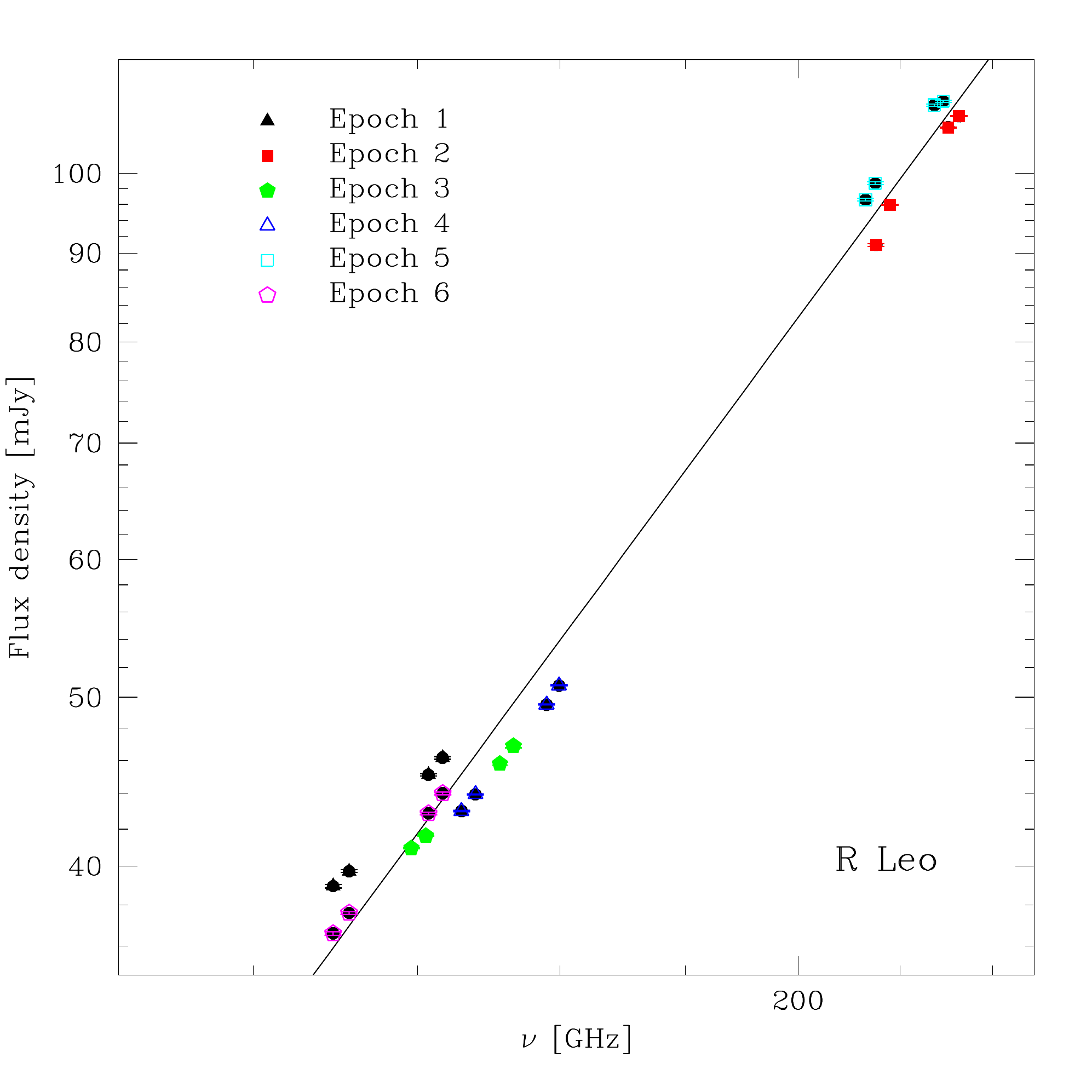}
 \end{minipage}
 \begin{minipage}[t]{0.5\textwidth}
      \includegraphics[width=\textwidth]{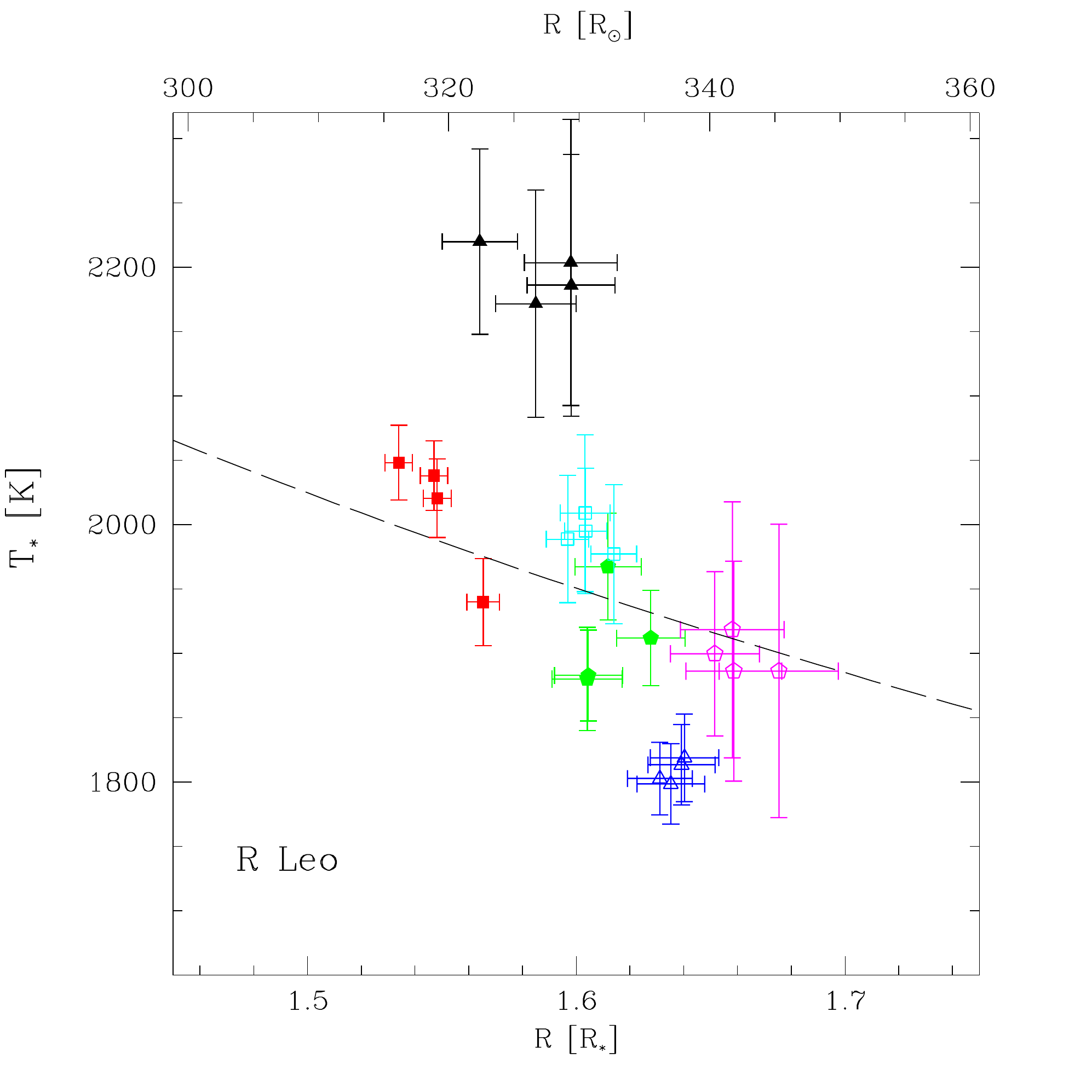}
 \end{minipage}  
\hfill
  \begin{minipage}[t]{0.5\textwidth}
      \includegraphics[width=\textwidth]{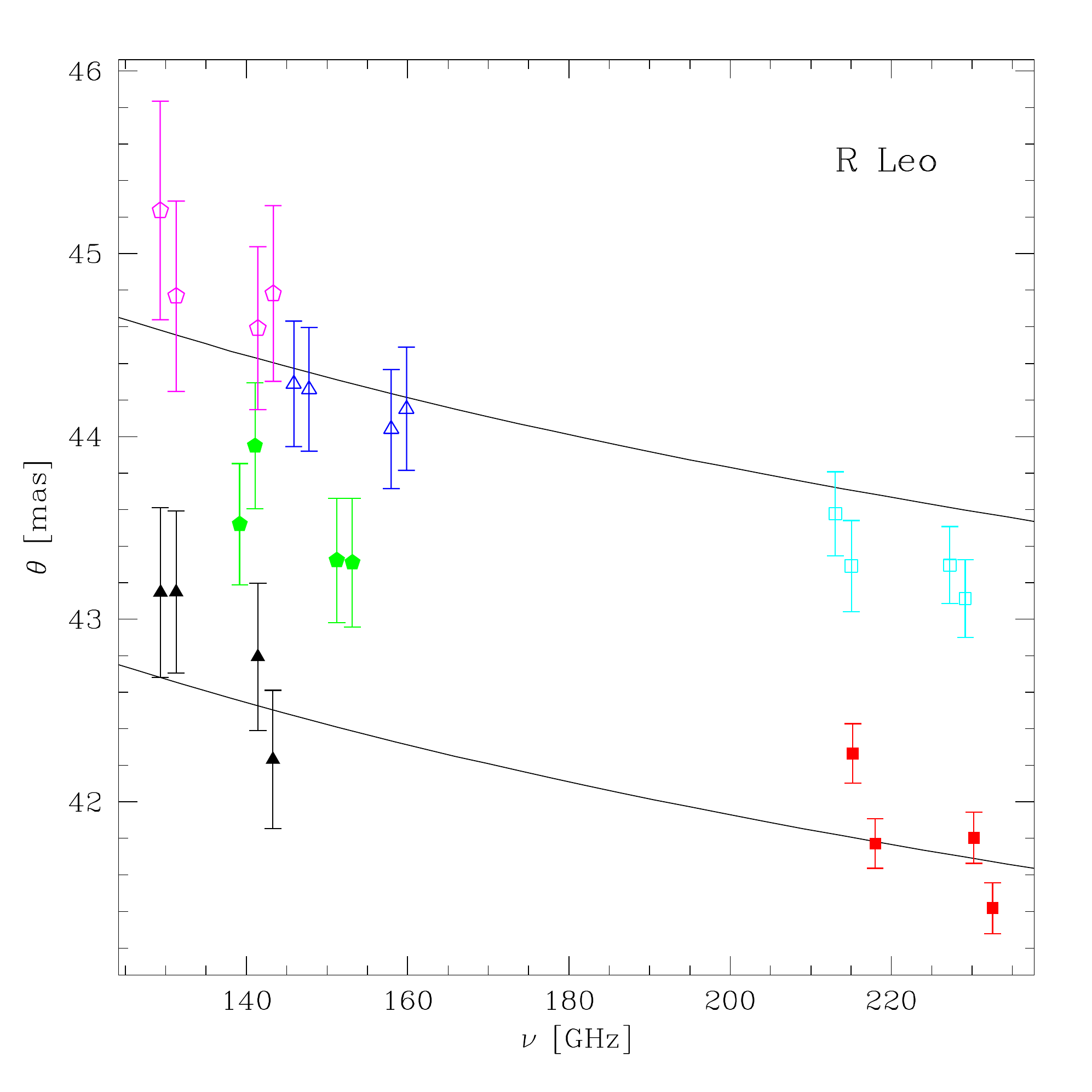}
 \end{minipage}
 \begin{minipage}[t]{0.5\textwidth}
      \includegraphics[width=\textwidth]{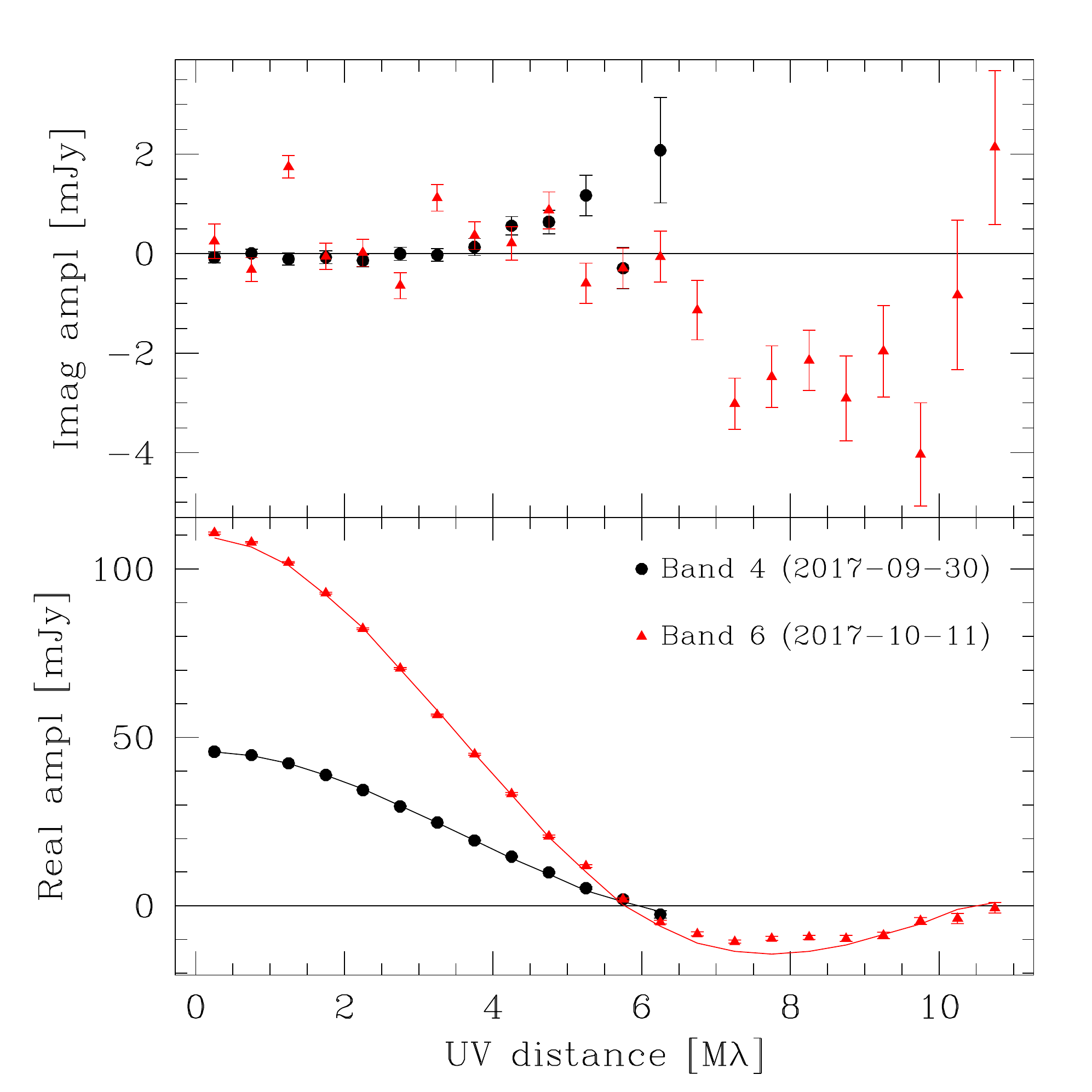}
 \end{minipage}  
  \caption{As Fig.~\ref{figMira} for R~Leo. The six different
     epochs are denoted by the symbols and colours as indicated in
     the top left panel. No fits are indicated for the individual
     epochs in the temperature vs. radius (top right) panel. The long-dashed line indicates the grey atmosphere temperature profile with a temperature at $R_*$ of
     $T=2850$~K. In the bottom left, the lines indicate the maximum
     and minimum power-law fit to the size vs. frequency relation
     when correcting each epoch for the observed expansion
     (\S~\ref{exp}). In the {\it uv} plot only two epochs, one for
     each observing band, are indicated for illustration.}
 \label{figRLeo}
\end{figure*}

Most of our analysis is performed using the {\it uv}-fitting code {\it
  uvmultifit} \citep{MartiVidal14} in a similar manner as described in
\citet{Vlemmings17}. We fit a uniform elliptical stellar disc for each
of the spws in our observations. This provides us with the flux density, major
axis, axis ratio, and position angle as a function of frequency and
time. For Mira~A, we performed the fits after its companion Mira B was
subtracted from the data. The results of our fits are listed in
Table~\ref{tab2} along with the observing date and stellar phase $\varphi$. In \citet{Matthews15}, an additional uncertainty was
added to the size measurements to account for residual phase errors on
the long baselines that arise from tropospheric phase variations. This
uncertainty depends on the baseline length and rms phase fluctuations
for the longest baselines. When using self-calibration, the phase
solutions are directly determined for each integration interval on
source. Hence, the rms phase fluctuations are significantly
reduced. Adopting the approach described in \citet{Matthews15}, we
conservatively find that a relative error term, with respect to the
measured size, of on average $\sim0.5\%$ in Band 4, $\sim0.2\%$ in
Band 6, and $\sim0.1\%$ in Band 7 should be added. The exact fractions
depend on the measured size and baseline lengths. The errors in
Table~\ref{tab2} reflect this added uncertainty.

The spectral indices provided in Table~\ref{tab2} are determined in
two ways, depending on whether different epochs were combined or
not. When epochs are combined for a fit, we include a $10\%$ absolute
flux density uncertainty for the different bands (except for R~Leo where we
adopt a $5\%$ uncertainty in Band 4). In that case, all spws in the
band are assigned the same fractional error and the spectral index and
its associated error are determined by a Monte Carlo scheme. For
several epochs we also present a spectral index fit using only the
four (or in some cases three) spectral windows. In that case, no
additional flux density errors are assumed besides the measurement
uncertainties. In neither case did we include the uncertainty in the
assumed spectral index of the flux calibrator. This error is of the
order of $0.05$ for our calibrators.

Although a uniform disc model provides good fits to all our
observations, residual structure clearly remains. This is
discussed in more detail in \S~\ref{hot}. While in most cases this
concerns small-scale structure, we found that for a number of spws,
particularly those in Band 7 of the Mira observations, excess flux at
short baselines affected the fits. This excess can be seen in the
  bottom right panel of
Fig.~\ref{figMira} for baseline lengths below $\sim2$~M$\lambda$. For
the affected spws we restricted our fits to {\it uv} distances $>2$, or
$>3$~M$\lambda$ for Band 4 and for Bands 6 and 7, respectively. The
possible origin of the excess, particularly for Mira~A, is discussed
in \S~\ref{excess}.

The measured sizes and flux densities are used to obtain the source brightness
temperature using
\begin{equation}
T_{\rm b}={{1.97\cdot10^{-8}\cdot S_{\nu}\cdot c^2}\over{\nu^2\cdot q\cdot\theta_{\rm
      maj}^2}}~{\rm K}.
\end{equation}
In this equation $S_{\nu}$ is the stellar flux density in mJy, $\theta_{\rm maj}$ the
major axis in mas, $q$ the axis ratio ($\theta_{\rm min}/\theta_{\rm maj}$), $\nu$ the frequency in GHz, and
$c$ the light-speed in m~s$^{-1}$. The results for the flux density
versus frequency, size versus frequency, and temperature versus size relations
are shown in Figs ~\ref{figMira},~\ref{figRDor},~\ref{figWHya},
and~\ref{figRLeo}. These figures also show the quality of the {\it uv} fits.

\begin{figure*}
 \centering
 \includegraphics[width=15cm]{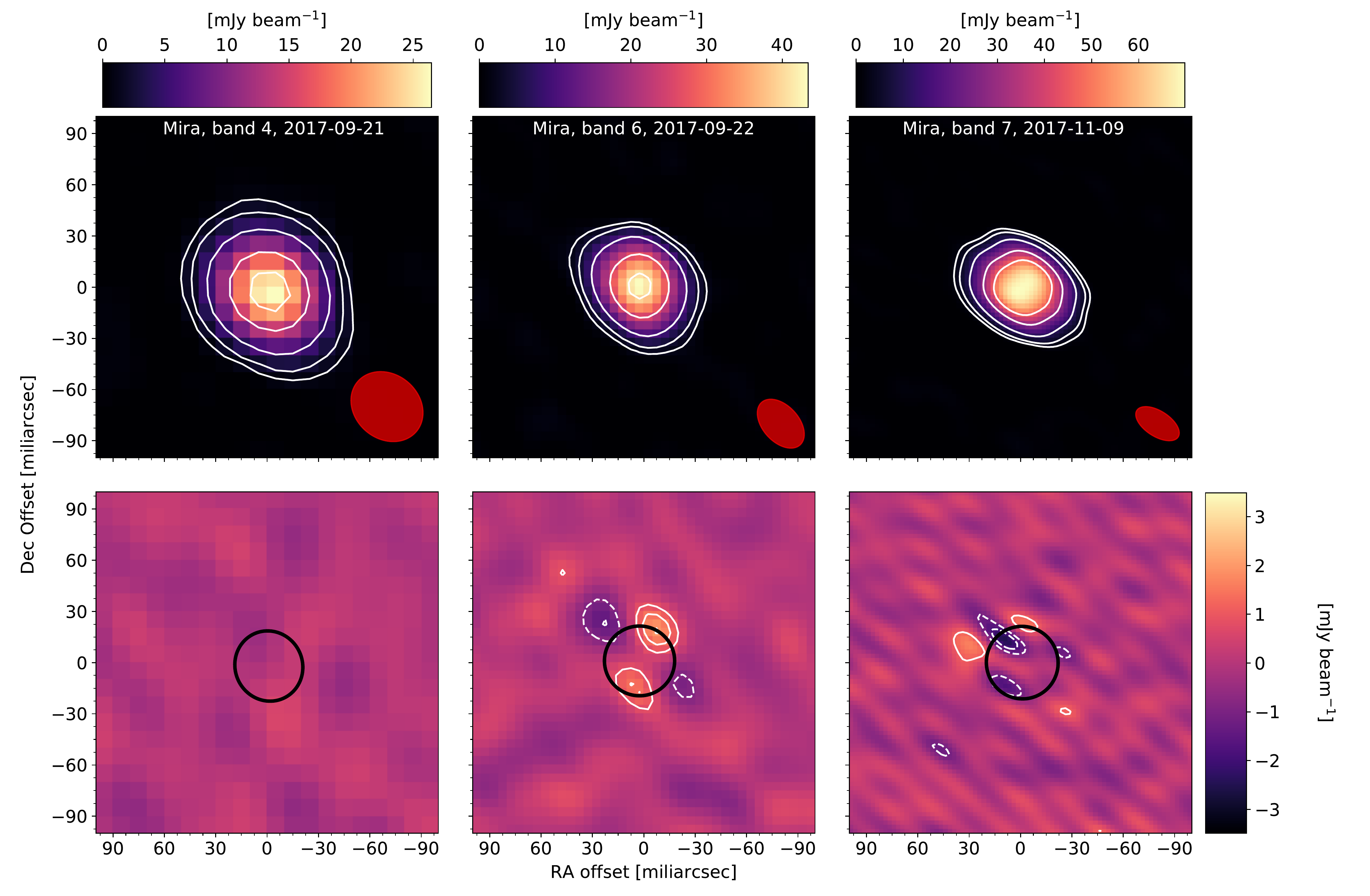}
 \caption{{\it Top row:} Uniformly weighted continuum maps of Mira~A in the three observed
   bands. The images serve as an illustration since the analysis is
   performed on the visibilities by {\it uv} fitting. The images are
   produced for a single spw, using the highest
     frequency spw in each observing band. Contours are drawn at $4, 8, 20, 50,$ and
   $80\sigma$, where $\sigma=0.28, 0.50,$ and $0.58$~mJy~beam$^{-1}$
   from left to right. The beam sizes, from left to right, are
   $45\times38$, $34\times21$, and $29\times15$~mas. {\it Bottom row:} Residual images after subtracting the
   best-fit uniform disc model presented in Table~\ref{tab2}. The models are fitted for the
   spectral windows individually and the subsequent residual image is
   produced using all spectral windows. The beam sizes, from left to right, are
   $45\times37$, $33\times21$, and $27\times13$~mas. Contours are
   drawn at $-15, -10, -7, -5, -3, 3,
   5, 7, 10, $ and $15\sigma$, where $\sigma=0.21, 0.27,$ and $0.37$~mJy~beam$^{-1}$
   from left to right. The black ellipse indicates the size of
 the best-fit uniform stellar disc model for the spw presented in the
 top panel.}
       \label{Miracont}
 \end{figure*}

\begin{figure*}
 \centering
 \includegraphics[width=15cm]{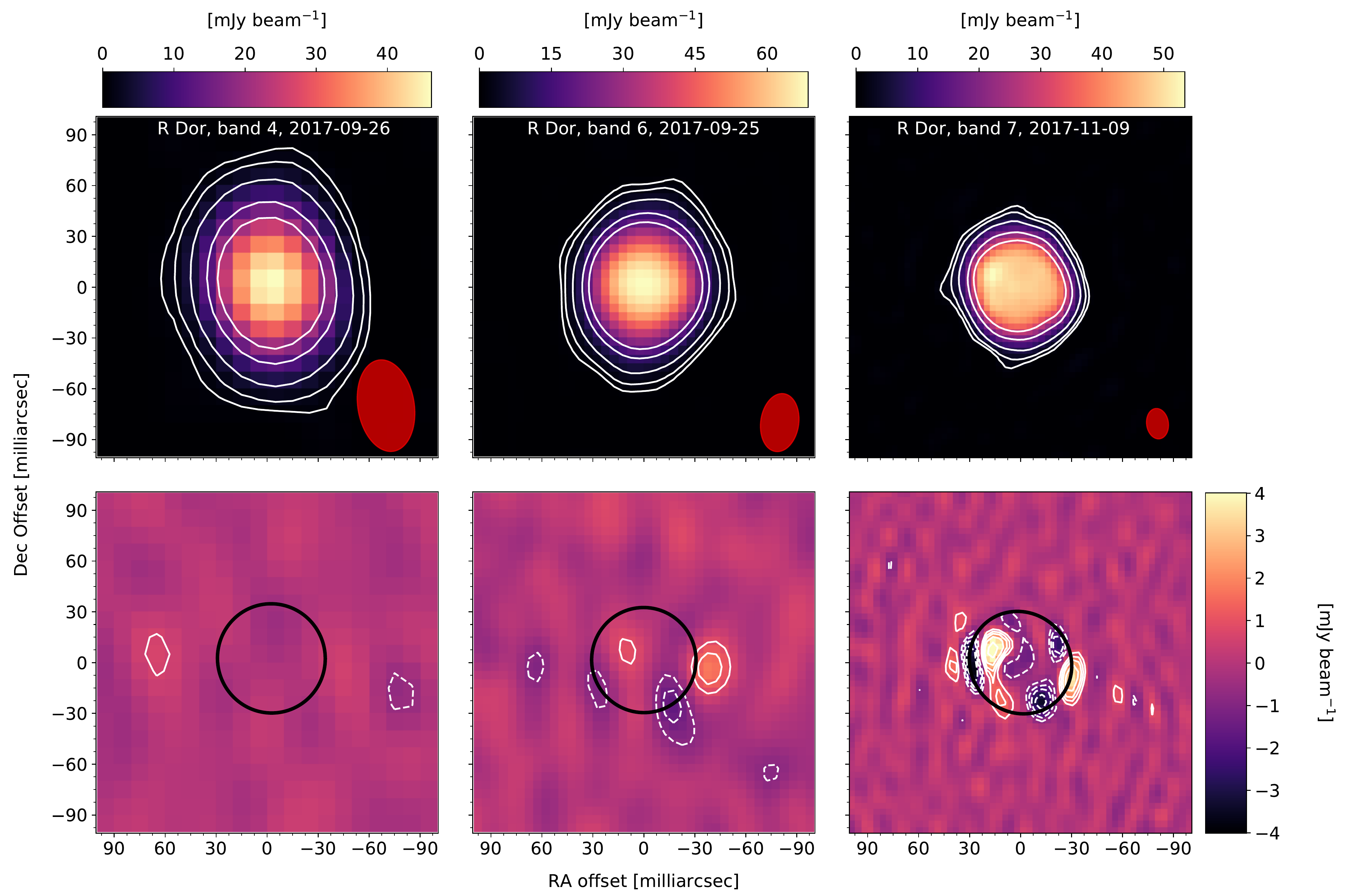}
 \caption{As Fig.\ref{Miracont} for R~Dor. Here $\sigma=0.25, 0.26,$ and $0.38$~mJy~beam$^{-1}$
   from left to right in the top panels, and $\sigma=0.17, 0.26,$ and $0.28$~mJy~beam$^{-1}$
   from left to right in the bottom panels. The beam sizes are
   $55\times33$, $34\times22$, and $18\times13$~mas on the top row
   and $52\times32$, $34\times22$, and $16\times11$~mas on the bottom
   row.}
       \label{RDorcont}
 \end{figure*}

\begin{figure*}
 \centering
 \includegraphics[width=10cm]{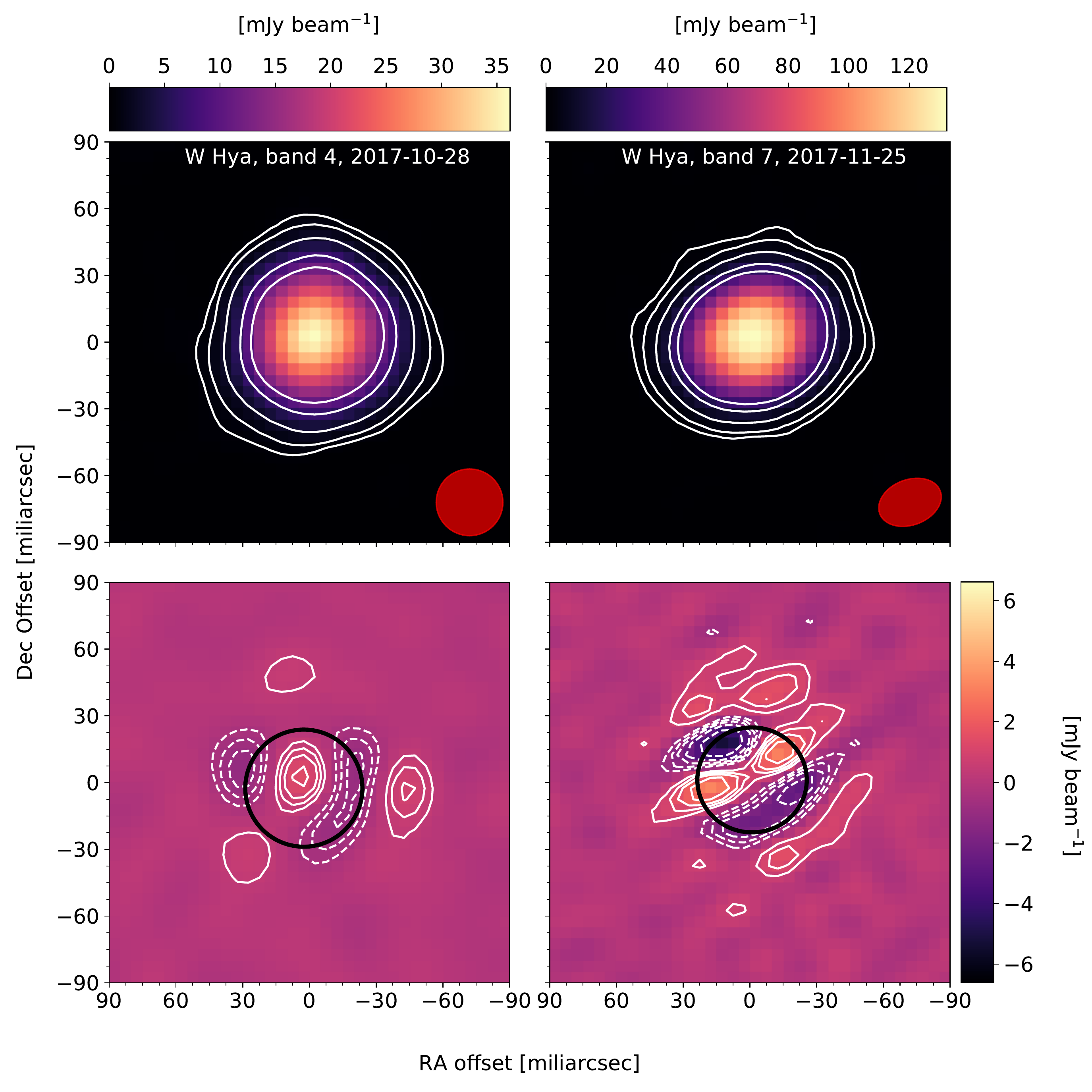}
 \caption{As Fig.\ref{Miracont} for the two observations of W~Hya, with $\sigma=0.14,$ and $0.33$~mJy~beam$^{-1}$
   from left to right in the top panels and $\sigma=0.12,$ and $0.22$~mJy~beam$^{-1}$
   from left to right in the bottom panels. The beam sizes are
   $30\times30$, and $29\times20$~mas on the top row
   and $30\times30$, and $27\times17$~mas on the bottom
   row.}
       \label{WHyacont}
 \end{figure*}

\begin{figure*}
 \centering
 \includegraphics[width=15cm]{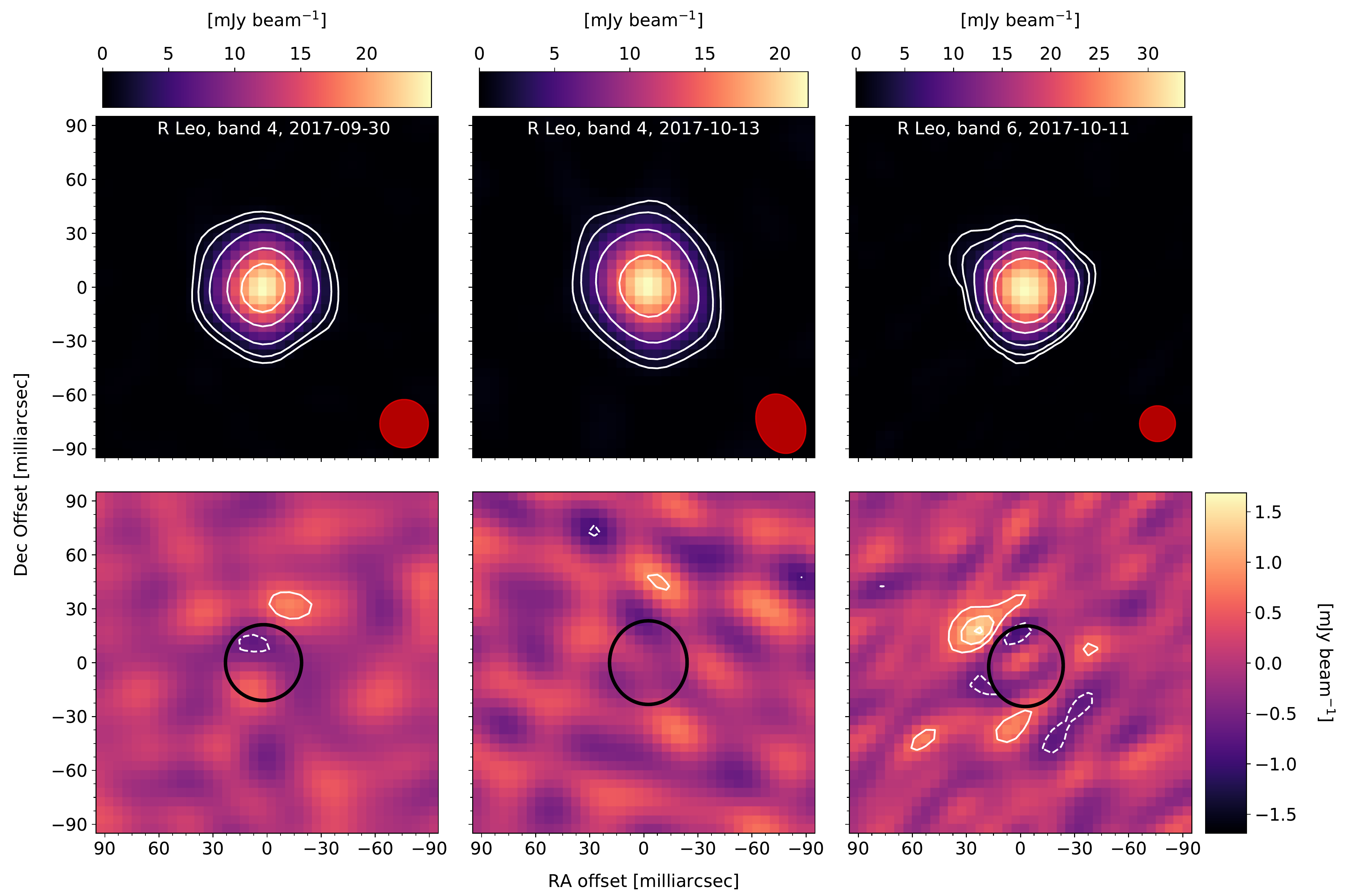}
 \caption{As Fig.\ref{Miracont} for R~Leo. Only three of the epochs
   are shown for illustration. The Band 6 epoch from 11 October was
   restored with a circular beam to highlight the northwest
   extension. In the top row, $\sigma=0.22, 0.30,$ and $0.25$~mJy~beam$^{-1}$
   from left to right. In the bottom row, $\sigma=0.19, 0.29,$ and $0.20$~mJy~beam$^{-1}$
   from left to right. The beam sizes are
   $27\times27$, $34\times26$, and $20\times20$~mas on the top row,
   and $31\times27$, $33\times25$, and $20\times20$~mas on the bottom
   row.}
       \label{RLeocont}
 \end{figure*}

For illustration, we also produced images for all of the observations
of Mira~A, R~Dor, and W~Hya, as well as three of the epochs for
R~Leo. These are shown in
Figs~\ref{Miracont},~\ref{RDorcont},~\ref{WHyacont},
and~\ref{RLeocont}. The images were produced using uniform weighting in
order to reach the highest resolution. Hence, the rms in the images is
significantly higher than the rms reached when using Briggs weighting
with a robust parameter of $+0.5$ (presented in
Table~\ref{tab1}). The figures also show the residuals after
subtracting the best-fit stellar disc model from the data in each spw.


\section{Discussion}

\subsection{Parameterised model atmospheres}
\label{models}

We constructed a
simple parameterised model atmosphere based on the radio-photosphere
model of RM97 to derive rough estimates for the physical conditions in the
extended atmospheres probed by our ALMA observations.  Since the density profile in the AGB atmosphere is
significantly affected by pulsations and convective motions, we adopt
a basic power-law description for the molecular hydrogen number
density $(n_{\rm H_2})$,
\begin{equation}
n_{\rm H_2} (r) = n_{\rm H_2, 0}\times \bigg({{r}\over{R_{\rm
      *}}}\bigg)^{-\alpha}.
\label{eq1}
\end{equation}
In this case $n_{\rm H_2, 0}$ is the number density (in $cm^{-3}$) at the stellar
photosphere, $R_*$ is the photospheric radius, and $\alpha$ is the power-law index. For comparison,
an atmosphere in hydrostatic equilibrium can, in the region of
interest, be described with a power-law exponent
$\alpha\approx9$. In a dynamical atmosphere, the exact density
versus radius relation becomes complex and depends strongly on pulsation
cycle with regions displaying slopes steeper than in the hydrostatic
case, while others are significantly more shallow
\citep[e.g.][]{Hoefner03}.  When convective motions are included, the
average power-law exponent can also reach $\alpha\approx6$
\citep{Freytag17}. Density profiles with $\alpha<6$ are also
typically used to describe the molecular atmosphere
\citep[e.g.][]{Wong16, Khouri16b}. We adopt $\alpha=6$ for our
reference model.

For the temperature profile we either use the grey atmosphere profile
also adopted in RM97 written as
\begin{equation}
T^4(r) = T^4_{R_*}\times(1-\sqrt{1-R^2_*/r^2}), 
\end{equation}
or a power law given by\begin{equation}
T(r) = T_{R_*}\times \bigg({{r}\over{R_{\rm
      *}}}\bigg)^{-\beta}.
\label{eq2}
\end{equation}
In this equation $T_{R_*}$ is the temperature (in K) at the photosphere and $\beta$ the
power-law index.
Finally, we also approximate the ionisation profile with a power law
in temperature, i.e.,\begin{equation}
 X_{\rm i} = X_{\rm i,3000}\times \bigg({{T}\over{3000}}\bigg)^{\gamma}.
\label{eq3}
\end{equation}
In this equation $X_{\rm i}=n_e/n_{\rm H_2}$, where $n_e$ is the electron
  number density.  We fix the ionisation at $T=3000$~K to be
$X_{\rm i,3000}=7\times10^{-6}$, which would correspond to single
ionisation of K, Na, Ca, and Al when assuming solar abundance. For the
specific stellar models presented later, the power-law index
$\gamma=8$, which fits the range provided by the detailed model
calculations from RM97.  The exact adopted value of $R_*$ for our
sources (as discussed in \S~\ref{sample}) does not affect any of our
conclusions, and for different stellar radii the surface values of
density and temperature can be scaled accordingly.

With these parameters, we can calculate the frequency dependent
optical depth throughout the extended atmosphere by taking into
account the relevant opacity sources. As found in RM97, of the
possible processes that contribute to the opacity, the electron-neutral free-free emission dominates
in the relevant temperature and density regime for the
radio and (sub)millimetre wavelengths. Hence we only considered this
process. We did confirm that the electron-ion free-free emission
contributes at least an order of magnitude less opacity even in
the warmest, most ionised, and densest regions of our standard
model. We only consider interactions of the electrons with purely
atomic hydrogen in our calculations. This is in contrast to the study
in RM97 in which an equilibrium state was assumed between atomic and
molecular hydrogen. As was already noted in \citet{Bowen88} however,
the formation rate of molecular hydrogen in the shocked atmosphere of
an AGB star can be slow. The RM97 authors argued that for high densities
($>10^{12}$~cm$^{-3}$) in the extended atmosphere, the equilibrium
time is $<250$~days. Since this is less than a typical pulsation
period, RM97 assumed equilibrium conditions. However, for most of our sources, we
find densities that are $<10^{12}$~cm$^{-3}$ throughout a large part
of the atmosphere. Also, recent models and observations
\citep[e.g.][]{Khouri16a, Freytag17} have shown that the timescales
for significant changes due to convective motions and/or shocks can be
measured in days or weeks. Hence we assume that equilibrium conditions
are not reached. In case there is a significant molecular hydrogen
component after all, the effective opacity decreases by up to a factor
of five and our derived densities (or ionisation) would need to be
increased. To calculate the opacity, we use the RM97 third-order
polynomial fits to the absorption coefficients calculated by
\citet{DL66}. We note that another calculation and approximation is
given in \citet{Harper01}, which is consistent within $\sim10\%$
across the temperature range that we probe in the AGB atmosphere.

\begin{table}
\caption{Model atmosphere: {\it fixed} parameters}             
\label{model}      
\centering          
\begin{tabular}{l l}     
\hline\hline       
$n_{\rm H_2, 0}$ & $4\times10^{12}$~cm$^{-3}$ \\
$\alpha$ & $6$\\
\hline
$T_{R_*}$ & $2500$~K\\
$\beta$ & grey atmosphere\\
\hline
$X_{\rm i,3000}$ & $7\times10^{-6}$\\
$\gamma$ & $8$ \\
\hline \hline
\end{tabular}
\end{table}

With the calculated opacities, we determine brightness temperature
profiles using standard radiative transfer in the Rayleigh-Jeans
limit. From these profiles we derive the frequency dependent stellar
disc brightness temperature and size that can be directly compared
with the observations. We assume a uniform stellar disc with a radius
$R(\nu)$, which is the radius where the brightness temperature
  has dropped by $50\%$ in units of stellar radii $R_*$. The angular
  size, $\theta$, can be directly related to $R(\nu)$ by dividing it
  by the adopted angular stellar radius from Table~\ref{Tab:sample}.
We focus in particular on the frequency-size and size-temperature
relations as presented in
Figs.~\ref{figMira},~\ref{figRDor},~\ref{figWHya}, and~\ref{figRLeo}
for the observations. As the opacity is proportional to the density
(with exponent $2$) and ionisation (linearly), in addition to the
inverse quadratic dependence on frequency, it is clear that there are
degeneracies between the various parameters used to describe the
extended atmosphere. Using the above exponents and the power-law
components from Eqs~$1-5$, we see that the optical depth
$\tau_{ff}\propto r^{-(\gamma\beta+2\alpha)} \nu^{-2}$. Since the
temperature is best constrained by the observations at different
wavelengths, the ionisation and number density are most
degenerate. Although the power-law approximation is likely far from
reality, with shocks and pulsations producing density and temperature
profiles that cannot be described using simple formulae, a number of
conclusions can still be drawn from a small parameter study. A
coupling of a dynamic and/or convective atmosphere code with the full
continuum radiative transfer is beyond the scope of this paper.

\begin{figure*}[htb]
   \begin{minipage}[t]{0.95\textwidth}
      \includegraphics[width=\textwidth]{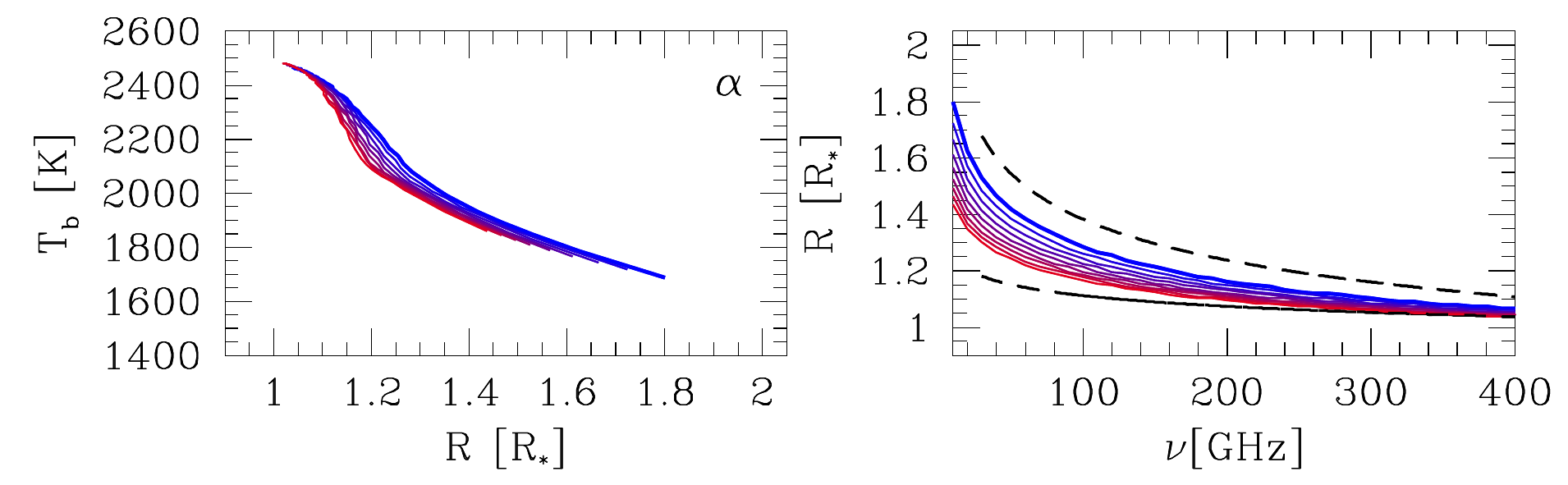}
 \end{minipage}
 \begin{minipage}[t]{0.95\textwidth}
      \includegraphics[width=\textwidth]{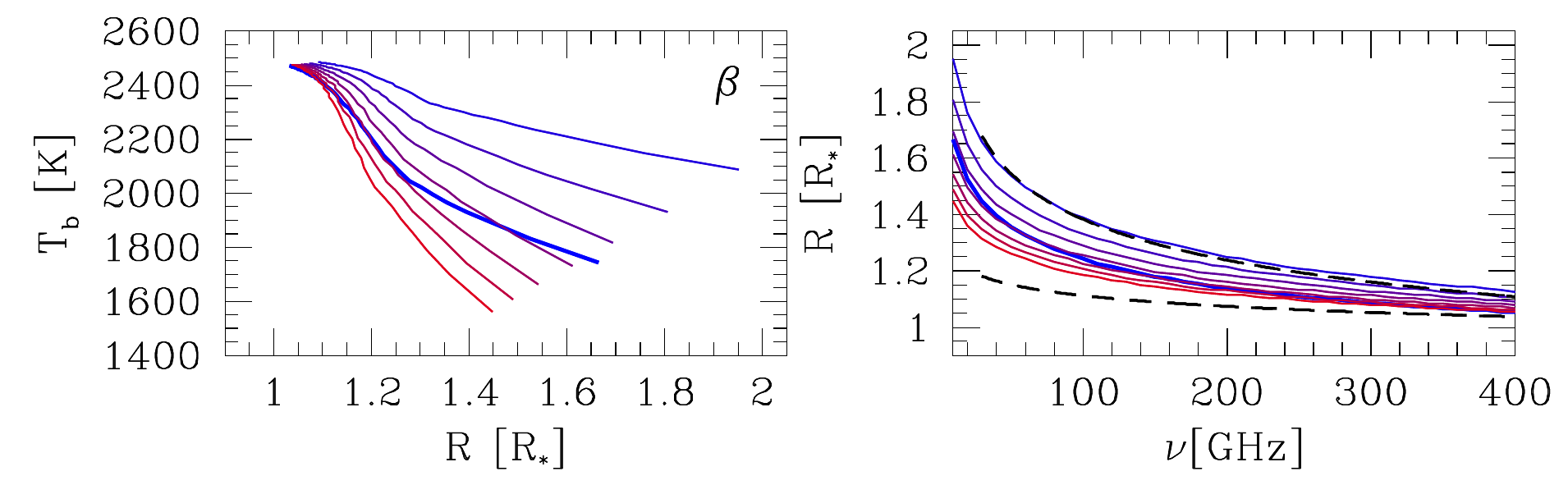}
 \end{minipage}  
\hfill
  \begin{minipage}[t]{0.95\textwidth}
      \includegraphics[width=\textwidth]{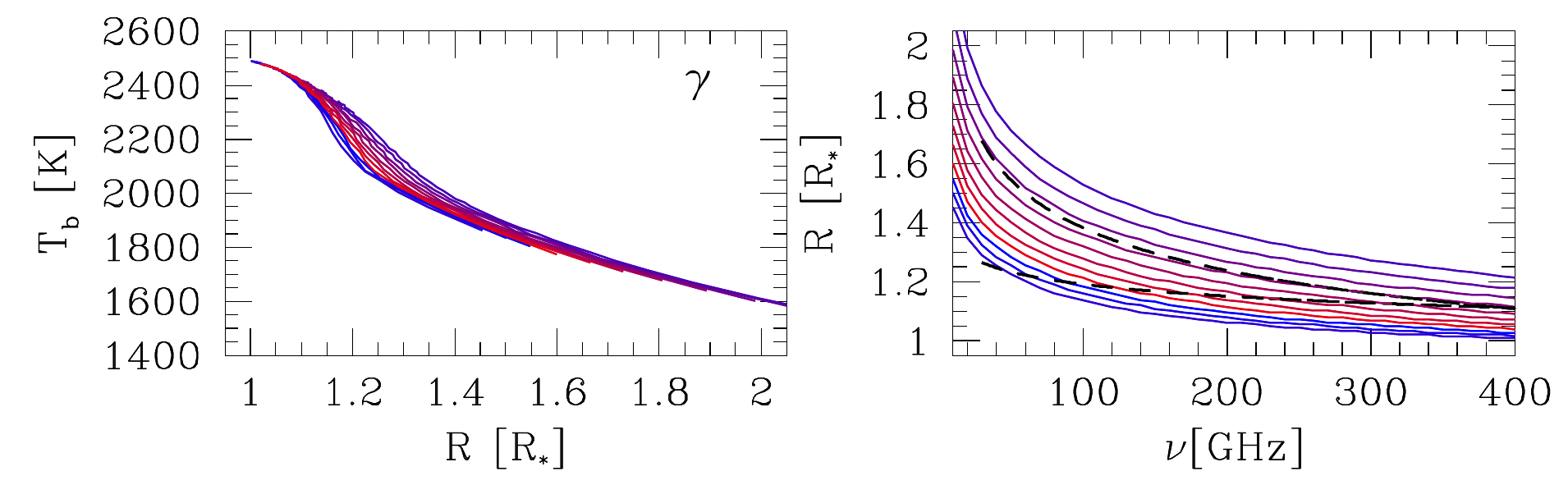}
 \end{minipage}
 \caption{Brightness temperature vs. size (left column) and size
   vs. frequency (right column)
relations for model atmospheres (see text) with
different power-law slopes for the density profile $(\alpha$, top),
temperature profile ($\beta$, middle), and
ionisation ($\gamma$, bottom). The two dashed
lines indicate the range of the observed slopes of the size vs. frequency
relation ($\eta$). Standard parameters are given in Table~\ref{model}. We vary $\alpha$ from $5$ (blue) to $9$ (red) in steps of
$0.5$. We vary $\beta$ from $0.3$ (blue) to $1.5$ (red)
in steps of $0.2$ and include the grey atmosphere model (dark
blue). We vary $\gamma$ from $2$ (blue) to $12$ (red)
in steps of $1$.}
 \label{mods}
\end{figure*}

In Fig.~\ref{mods}, we show the size versus frequency and brightness temperature
versus size relations for models with different $\alpha, \beta,$ and
$\gamma$. The fixed model parameters for the plotted models are given
in Table~\ref{model}. In the figures we also indicate the observed
range of slopes $(\eta)$ of the size versus frequency relation derived
from the observations. From this is it clear that the lowest of
  the derived values of
$\eta$ observed for Mira~A, R~Dor, and R~Leo cannot be reproduced for any but the steepest density profiles
($\alpha>9$) and for none of the temperature or ionisation
profiles. One of the most straightforward ways to match the shallow
size versus frequency relations observed is to include a transition
region with a sudden increase in opacity. Such a transition could
naturally be related to a shock travelling through the extended
atmosphere and mimics the density behaviour in dynamical atmopshere
models \citep[e.g.][]{Bowen88, Hoefner03} that have density variations up
to a factor of $\sim10$. We thus produce a model, which we describe as
a ``layer'' model that has a ten times increase in $\tau_{\rm ff}$ at
a single ``shock'' radius ($R_s$) to investigate if this could reproduce the
observations. Including such an opacity increase specifically changes the
stellar profile by introducing a region that is sufficiently optically
thick so that there is no noticeable change of size with increasing
frequency (within the observed frequency range). It can also produce
limb brightening effects or ring-like structures in the residual maps
once the layer has moved further out. As a result, the size determined
using a uniform stellar disc increases, thereby producing a sharp
drop in the temperature versus size relation and a leveling off of the
size versus frequency relation. Ring-like structures related to this
effect have possibly been seen around Betelguese \citep{OGorman17} and
IRC+10216 \citep{Matthews18}.

\begin{figure*}[htb]
   \begin{minipage}[t]{0.5\textwidth}
      \includegraphics[width=\textwidth]{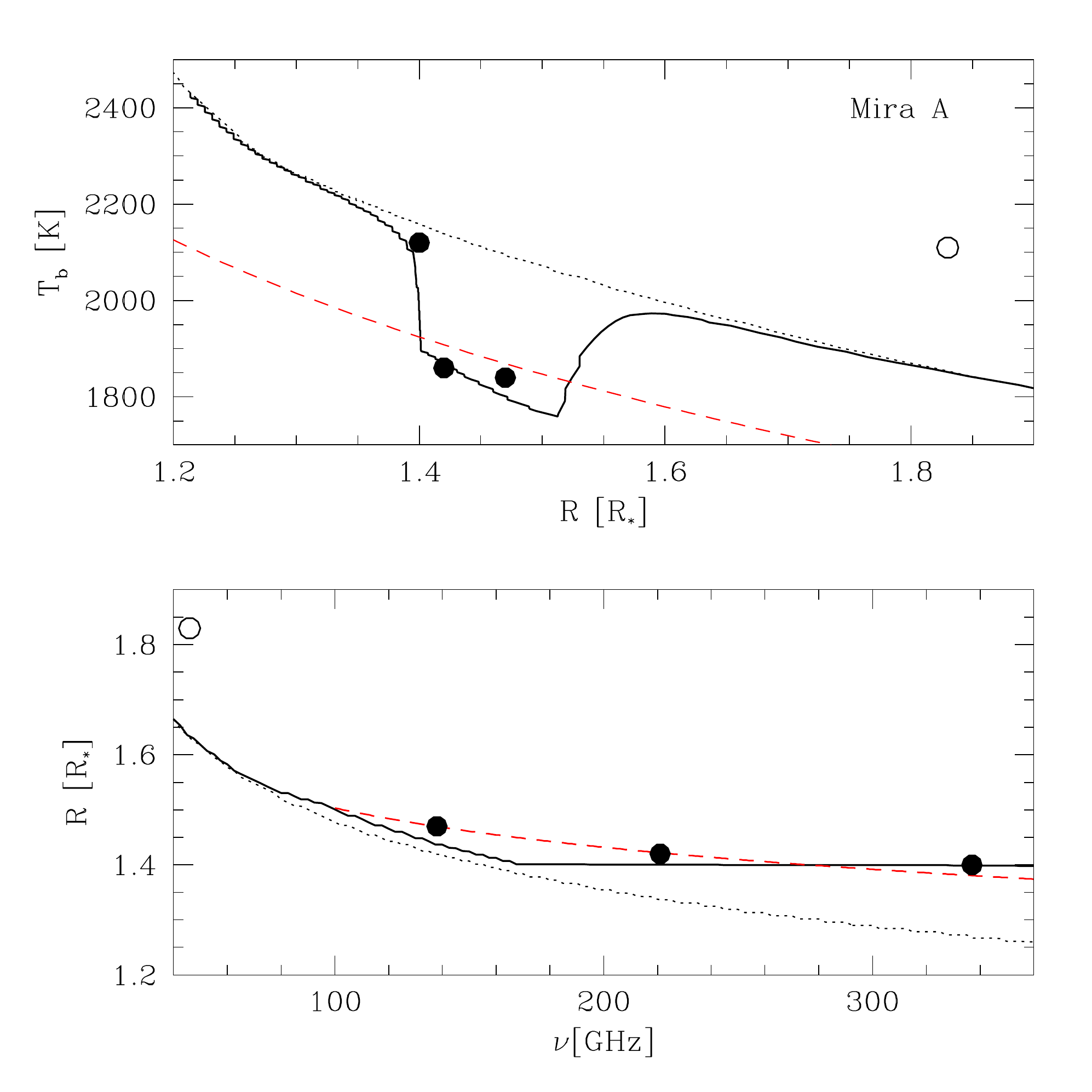}
 \end{minipage}
 \begin{minipage}[t]{0.5\textwidth}
      \includegraphics[width=\textwidth]{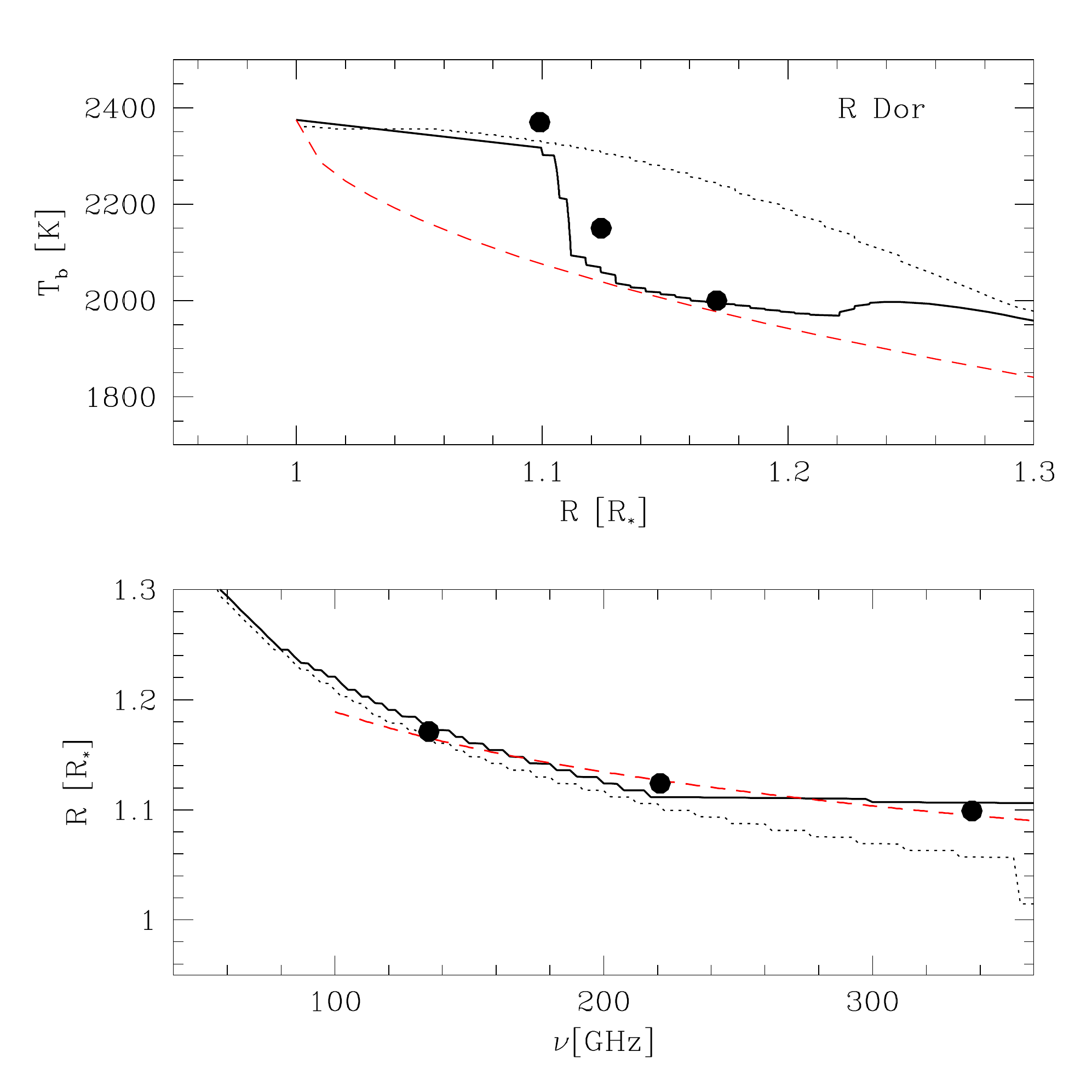}
 \end{minipage}  
\hfill
  \begin{minipage}[t]{0.5\textwidth}
      \includegraphics[width=\textwidth]{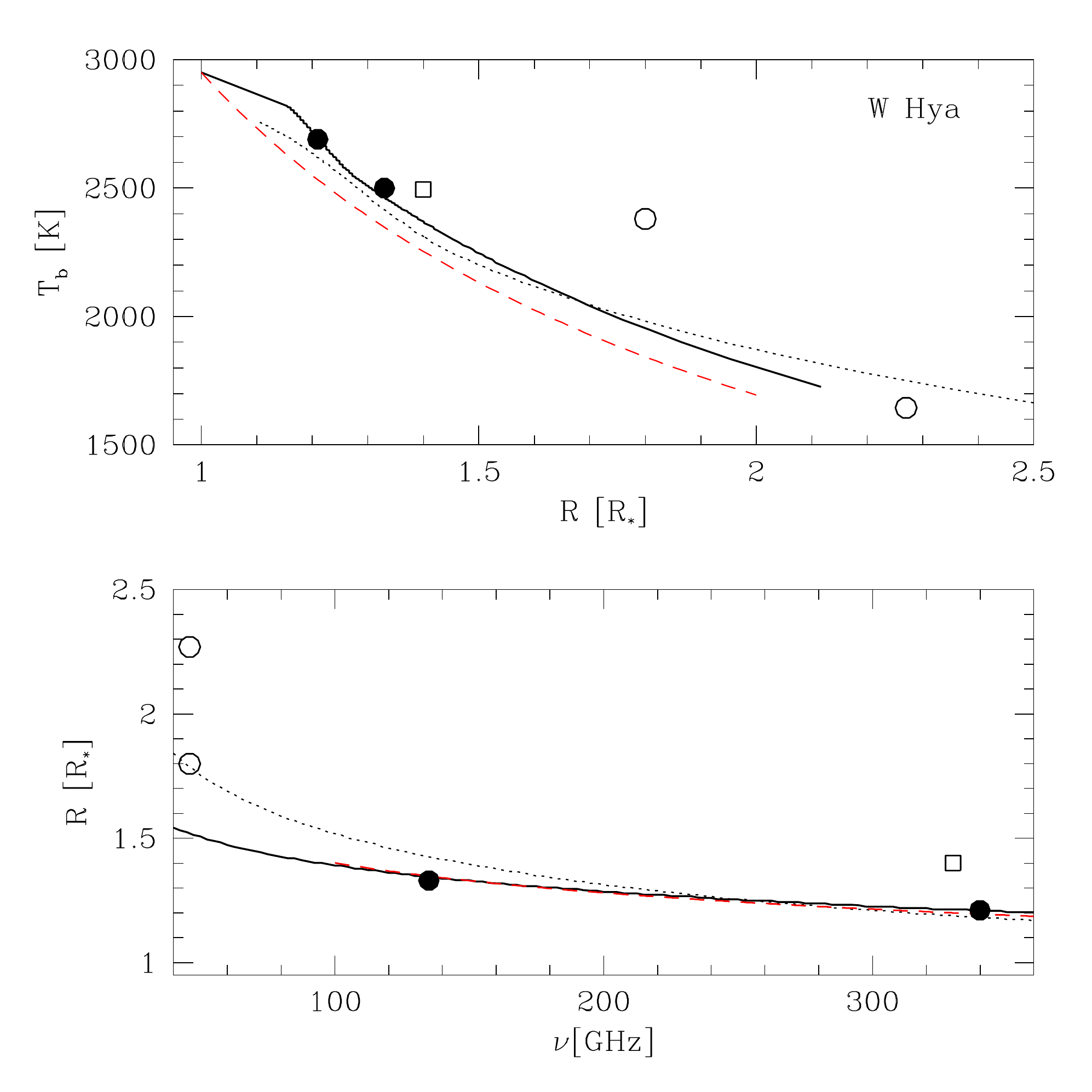}
 \end{minipage}
 \begin{minipage}[t]{0.5\textwidth}
      \includegraphics[width=\textwidth]{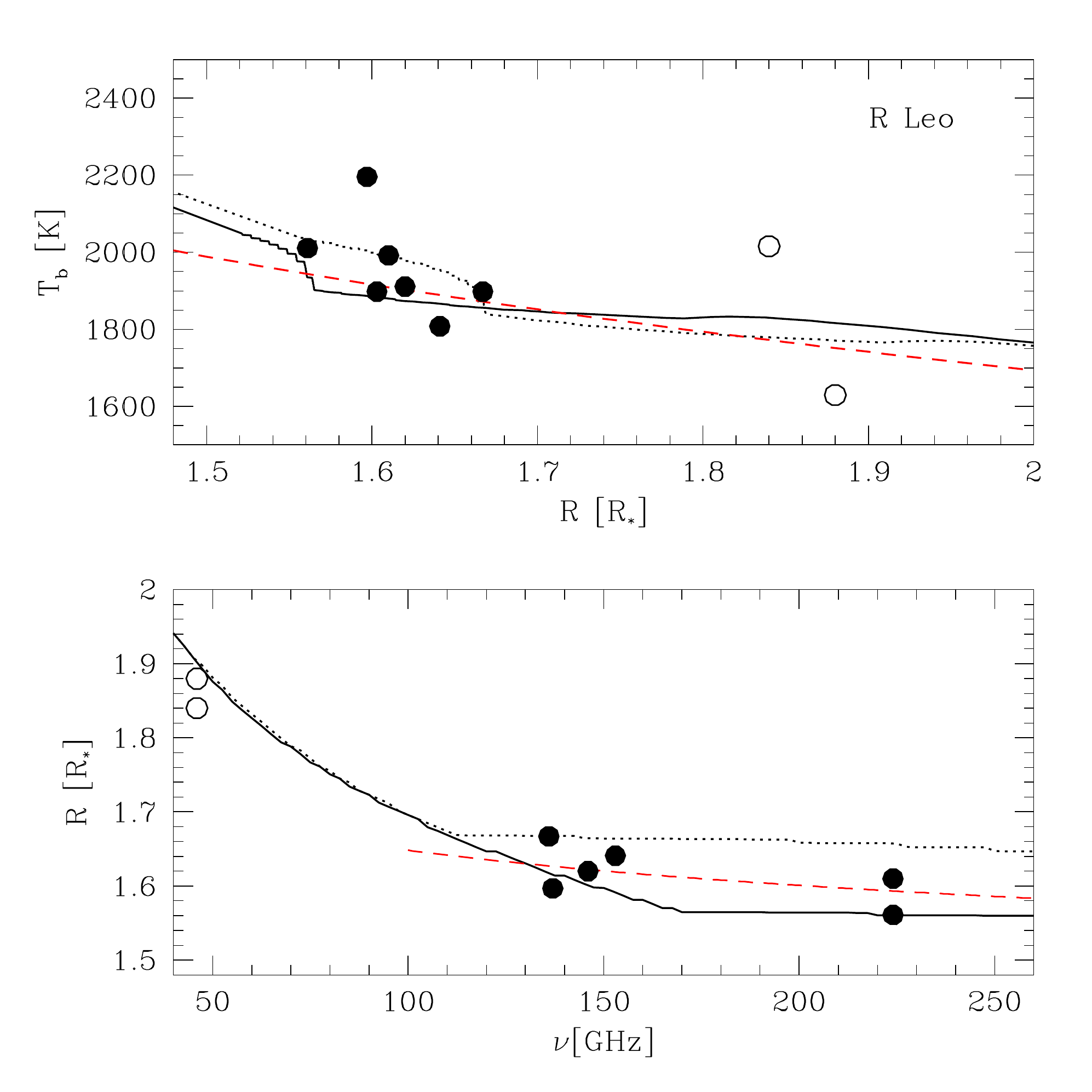}
 \end{minipage}  
 \caption{Extended atmosphere models for Mira~A (top left), R~Dor
   (top right), W~Hya (bottom left), and R~Leo (bottom right) that can
   reproduce our observations. For each star the top panel shows the
   temperature vs. radius relation and the bottom panel shows the size
   vs. frequency relation. The solid circles are the measurements
   presented in this paper, and are taken as an average of all spws
   per observations. The open circles are 46~GHz radio observations
   \citep{Matthews15, Matthews18} shown for comparison. The radio
   observations were taken several years before the ALMA
   observations. In the panels for W~Hya, the open square indicates the
   ALMA observations taken approximately 2~years earlier \citep{Vlemmings17}. The thick solid lines are the models described in the
   text. In the top panels for each star, the red dashed line represents the assumed
   temperature profile. In the bottom panels for each star, the red dashed lines
   represent the size vs. frequency power-law relation fitted to the
   observations. For Mira~A and R~Dor, the dotted line represents a
   model without a high-opacity layer. For W~Hya the dotted line
   represents the grey atmosphere model with $n_{H_2}\propto r^{-3}$
   that comes closest to describing the radio observations. For R~Leo
   the dotted line represents the model for which the layer location
   has shifted (see text).}
 \label{modstars}
\end{figure*}

We discuss the modelling results for the
four stars separately. We adopted the standard ionisation description
for all stars. We note that we treat the different epochs presented in
this paper for all stars except R~Leo as sufficiently simultaneous in
this analysis. The results are shown in Fig.~\ref{modstars}.

\subsubsection{Mira~A}

The slope of the size versus frequency relation for Mira~A is shallow,
where $\eta=-0.087\pm0.006$ is determined from the Band 4 and 6
observations, and $\eta=-0.053\pm0.002$ between the Band 7
observations taken $\sim1.5$~month later and the Band 4 observations.
The observations can be reproduced by
a typical density profile with $n_{H_2,0}=1\times10^{13}$~cm$^{-3}$
and $\alpha=6$. For the temperature, a grey atmosphere with
$T_{R_*}=2800$~K is consistent with the measured brightness
temperatures. Additionally, the low $\eta$ and apparent abrupt drop of
$T_b$ between the Band 6 and 7 observations requires a sudden increase (moving inwards) in opacity by at least a factor of 10 at
$R_s=1.4~R_*$  in our model
framework. We do not cover a sufficient range of radii to
determine the thickness of the region with elevated
opacity. Considering the relation between opacity and the modelling
parameters, the opacity change can be produced by an increase of the
density by at least a factor of three or a change of ionisation by a
factor of $10$ or a combination of these. A significant change in
temperature would have affected the brightness temperature versus size relation. To
indicate how our model compares with radio-wavelength observations at
46~GHz \citep{Matthews15}, we also present the average of the major and
minor axis of these observations in the figure even though the radio
epoch was three years before the ALMA observations. We find that our
model underestimates the brightness temperature measured at 46~GHz by almost
$20\%$ while it underestimates the radius by $10\%$. Considering that the
temperature uncertainty of the radio observations is $\sim20\%$ and
the ellipticity is almost $18\%$, the model could be consistent. The
size measurements of the previous ALMA observations at 96 and 230~GHz
are also consistent but the derived brightness temperatures are
$>20\%$ higher \citep{Vlemmings15, Matthews15}.

The layer of enhanced opacity we derive for Mira~A is likely related
to shocks originating from stellar pulsations and/or convection.
Since we do not have sufficient observing epochs spread over a few
weeks timescale taken at similar frequencies, we are unable to
determine a motion of the layer. Alternatively, it might not be an
actual spherical layer, as we observe asymmetries such as a mild
elliptical shape and compact structures in our residual maps. Our
models are not yet able to determine the effect of such
structures. The derived density and temperature relations should thus
be seen as rough indications only. Still, deviations from the
densities we find by a factor of three or more would, in our
framework, produce significantly different stellar sizes and are thus
unlikely unless compensated by differences in ionisation or another
opacity source.

\subsubsection{R~Dor}

The sizes of R~Dor observed at the different frequencies are, in
stellar radii, the smallest of our sample. Similar to Mira~A, the
change of size with frequency is low ($\eta=-0.078\pm0.001$ between
Band 4 and Band 6, and $\eta=-0.059\pm0.013$ in Band 7). A combination
of a fast drop in temperature and the low $\eta$ also implies the
existence of a layer with increased opacity around R~Dor. Since the
measured size is small, we can reproduce the profiles with a lower
density of $n_{H_2,0}=4\times10^{12}$~cm$^{-3}$ and $\alpha=6$. The
temperature is well reproduced by a grey atmosphere with
$T_{R_*}=2375$~K. The opacity enhancement (by a factor of $10$) around
R~Dor occurs at $R_s=1.11~R_*$. For R~Dor, the ellipticity increases markedly
towards the higher frequencies in Band 7 and also several clear
residual emissions occur both towards the star and in the
limbs. Although the similar caveats should be considered as for
Mira~A, the observations imply that the layer is related to the
asymmetric structures that are likely (convective)
shocks. Alternatively or additionally, the increased opacity so close
to the photosphere could arise owing to the presence of a higher
ionisation region related to a possible chromosphere. No long
wavelength observations of R~Dor exist for comparison because of its
low declination.

\subsubsection{W~Hya}

W~Hya is the only star in our sample that displays a slope $\eta$ that
can be reproduced without adding a region with enhanced opacity
($\eta=-0.160\pm0.007$ in Band 4, and $\eta=-0.104\pm0.003$ in Band
7).  For this, we need a density
profile with $n_{H_2,0}=4.5\times10^{12}$~cm$^{-3}$ and
$\alpha=6$. A temperature profile with $T_{R_*}=2950$~K
and $\beta=0.8$ produces a slightly better correspondence with the
data than a grey body atmosphere model. However, a grey body
atmosphere with $n_{H_2,0}=3\times10^{12}$~cm$^{-3}$ and
$\alpha=3$ is better able to reproduce the radio observations as was
already noted for W~Hya by RM97. This likely reflects the complex
nature of the AGB extended atmosphere where a simple single power law
is not sufficient to characterise the density and temperature. W~Hya also shows
significant residuals from a simple stellar disc model, both in the
observations presented in this work and the previous ALMA observations
\citep{Vlemmings17}.

\subsubsection{R~Leo}

R~Leo presents a complex case. As we discuss in \S~\ref{exp}, a
clear expansion is seen. This points to an outward motion of a shock
wave. When adjusting each epoch for the fitted outward motion of
$V\approx11$~km~s$^{-1}$ and fitting a single power law to the adjusted size
versus frequency relation, we find $\eta=-0.042\pm0.015$. We can
reproduce the range of observed sizes and temperatures by introducing
an increased opacity layer as shown in Fig.~\ref{modstars}. The layer moves
from $1.57~R_*$ to $1.66~R_*$ and the underlying density profile has
$n_{H_2,0}=1\times10^{13}$~cm$^{-3}$ and $\alpha=5.0$. The temperature
is described by a grey atmosphere model with $T_{R_*}=2800$~K. Since
it appears that at least part of the expansion of R~Leo is asymmetric, with
significant residuals from a uniform stellar disc towards the
north-east, the apparent layer might in reality be a directional
shock wave propagating through the extended atmosphere.

\subsection{Expansion of R~Leo}
\label{exp}

\begin{figure}
\centering
\includegraphics[width=0.47\textwidth]{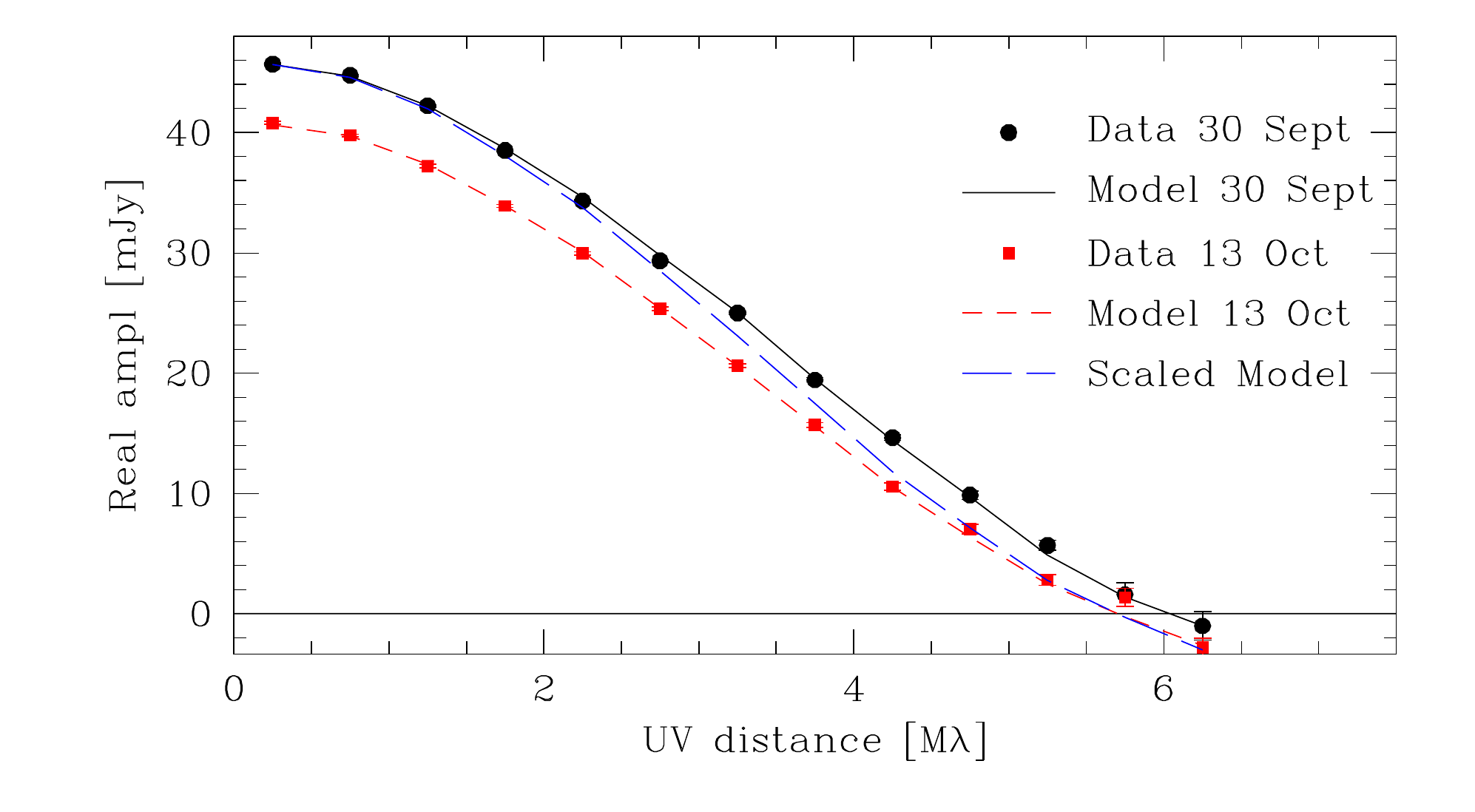}
\caption{Comparison of the real part of the {\it uv} visibilities
 between R~Leo Band 4 observations taken on 2017-09-30 (black solid
 line and circles) and
 2017-10-13 (red dashed line and squares). In most cases the error
 bars are less than the symbol size. The lines represent the
 uniform disc model, where the long-dashed blue line corresponds to
 the best-fit model of the 2017-10-13 observations scaled to the flux density
 of the 2017-09-30 observations. Although the difference in flux density is
 most likely predominantly
 caused by the uncertainty in absolute flux calibration, the
 long-dashed blue line shows that the change in radius of the
 model is robustly determined. }
     \label{RLeo_uvcomp}
\end{figure}

\begin{figure}
\centering
\includegraphics[width=0.47\textwidth]{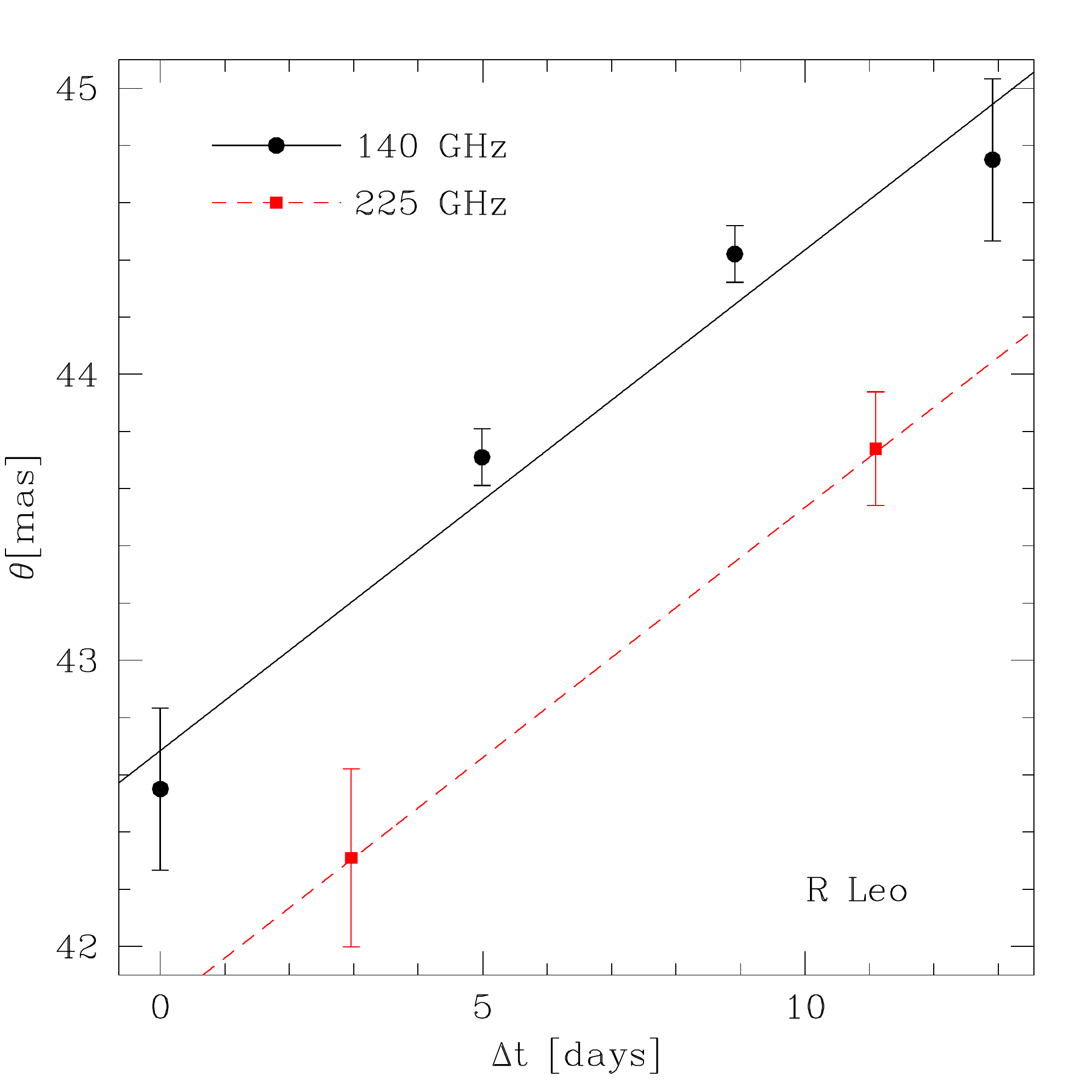}
\caption{Average diameter $\theta$ vs. observing time after the
  first epoch (2017-09-30) for the star R~Leo. The black circles indicate the radius at
  140~GHz in Band 4 and the red squares are the fitted radii at
  225~GHz in Band 6. The black solid line is the error weighted fitted
  expansion of the optically thick surface. The expansion velocity
  corresponds to $10.6\pm1.4$~km~s$^{-1}$ (assuming a distance of
  $71$~pc). The red dashed line is the fit to the $140$~GHz radii
  shifted by a time lag of $-5$ days.}
     \label{RLeoExp}
\end{figure}

ALMA observed R Leo within two weeks on four and two occasions in Band
4 and Band 6, respectively, at fairly similar frequencies in each
given band. We are thus able to investigate its time variable behaviour
in detail. Looking at the flux density variations, we find differences of on
average $5\%$. These can mostly be attributed to absolute flux
calibration uncertainties, even though the same flux calibrator was
used. However, we also note a clear trend of increasing radius and
ellipticity with time.  In Fig.~\ref{RLeo_uvcomp} we show a comparison
of the {\it uv} data and model fits for the first and last observing epoch
in Band 4. The fit indicates that the size change is robustly
determined. In Fig.~\ref{RLeoExp} we display the measured size against
the time after the first observations for both Band 4 and Band 6. In
this plot we adjust all measurements to 140 and 225~GHz for the
Band 4 and Band 6 data, respectively, using the individually fitted
slopes to the frequency-size relations. We can fit a linear expansion
of $0.17\pm0.02$~mas~day$^{-1}$ to the Band 4 data and find that this
is consistent with the size measurements during the two epochs at
225~GHz. These measurements imply that the continuum surface seen in
the millimetre wavelength observations is expanding at
$10.6\pm1.4$~km~s$^{-1}$ when assuming a distance to R~Leo of 71~pc. The
fact that the 140 and 225~GHz surfaces undergo a similar expansion
implies a bulk motion of material that encompasses almost $0.2~R_*$. A
motion of the continuum surface can occur because of changes in
density, ionisation and/or temperature. The temperature in the region
that we are probing does not seem to change by more than $\pm5\%$
(most likely due to absolute flux calibration uncertainties).
Such a variation would produce a change in ionisation of $\sim50\%$ and
a similar change in opacity. According to our models for R~Leo, this
would result in a change of measured radius, at 140~GHz, of
$\sim0.05~R_*$. This assumes the temperature changes throughout the
entire envelope which requires radiative heating, which is
unlikely. The radius change would be significantly less if the
temperature change is only local. Hence, the motion that we are seeing
is likely due to density or ionisation changes (by factors $>2$ or
$>4$, respectively).

The measured velocity matches the velocity dispersion seen in the
gravitationally bound molecular gas around a number of stars
\citep[e.g.][]{Hinkle97,Wong16, Vlemmings17, Vlemmings18}. Since the star also increases
its ellipticity, the motion is not spherically symmetric. The
expansion velocity taking the major axis values alone would correspond
to $12\pm1$~km~s$^{-1}$ and that of the minor axis to
$9\pm1$~km~s$^{-1}$.  Note that if the movement occurs mostly in only
one direction and not equally in opposite directions (which is unlikely
if it is a result of convective motions), the inferred velocities are
up to a factor of two higher. In any case, the velocity is less than
the escape velocity, which is $\sim34$~km~s$^{-1}$ at $1.6~R_*$ for
a star with $M=1$~M$_\odot$. The velocity exceeds the local sound
speed in the atmosphere, which is $\sim5$~km~s$^{-1}$. It is also
larger than the velocities of $\lesssim8$~km~s$^{-1}$ found
at $R\sim1.6~R_*$ in current convection models \citep{Freytag17}.

\subsection{Asymmetries and hotspots}
\label{hot}

Previous observations of AGB stars at radio and sub-mm/mm wavelengths have
revealed a variety of asymmetries and hotspots in the extended
atmospheres. Radio observations at different epochs have shown
significant ellipticity, brightness asymmetries and non-uniform
surfaces \citep[e.g. RM97][]{RM07, Matthews18}. 
ALMA observations of Mira~A and W~Hya also revealed a mottled surface
structure at sub-mm/mm wavelengths \citep{Vlemmings15, Matthews15, Vlemmings17}. The most
prominent of these was the detection of a hotspot with $T_{\rm b}>50000$~K
on the surface of W~Hya. As can be seen in the residuals from a
uniform stellar disc model and the {\it uv} model fits for the sources
presented in this work (Figs~\ref{Miracont},~\ref{RDorcont},~\ref{WHyacont},
and~\ref{RLeocont}), all stars display asymmetries and/or hotspots to
various degree. In some cases, negative residuals are seen that
  are similar in strength to the positive features. This is a natural
  result when determining the residuals after subtracting a best-fit uniform-disc model, which minimises the intergrated flux
  across the disc.

{\bf Mira~A:} Our observations of Mira~A indicate that, at the current
epoch, the star is relatively circular with a marginally significant
ellipticity of a few percent seen in the Band 6 data. This is in
contrast with previous ALMA observations that indicate an ellipticity
between $10-20\%$ \citep{Vlemmings15, Matthews15}.
While the previous observations had significant residuals that could
indicate a strong hotspot, the current data only have a weak spotted
structure, predominantly in the limb of the disc. This could be related
to a (non-uniform) enhanced opacity layer, such as included in the
atmosphere models, with changes in density or ionisation by a factor of
$\sim3$ or $\sim10$, respectively.

{\bf R~Dor:} R~Dor shows strong residual hotspots, especially in the
Band 7 observations. At those frequencies we can determine a lower
limit to $T_{\rm b}$ for both brightest hotspot of $>3300$~K above the
average stellar disc temperature. The actual temperature could be
significantly higher if the hotspot sizes are smaller than the formal
limit of $<6\times3$~mas obtained from a fit to the residual image and
{\it uv} data. The brightness variation across the stellar surface appears
to correlate with structure seen at visible wavelengths \citep[][and
in prep.]{Khouri18b}. In a comparison between the Band 6 and 7
observations, there also appears to be a correspondence between the
structure seen at both wavelengths. The two frequencies probe a
slightly different region of $\sim0.03$~R$_*$ in depth. If the
structures are indeed the same, they would thus have a vertical extent
of at least $8\times10^{11}$~cm ($\sim0.05$~au). Since R~Dor appears
to be the star with the smallest (sub)millimetre size compared to its
stellar radius as determined from infrared observations, we are
potentially probing very close to a chromosphere if it is
present. Notably, the ellipticity of R~Dor increases towards the
shorter wavelengths in Band 7, with an ellipticity up to $\sim6\pm1\%$
at a position angle of $44\pm2^\circ$. There is no correspondence
between this position angle and the apparent rotation axis
\citep{Vlemmings18}.

{\bf W~Hya:} As was found previously \citep{Vlemmings17}, W~Hya shows
strong deviations from a uniform disc. Hotspots and negative
patches are seen in both the Bands 4 and 7 observations. As
  discussed above, in the
event of strong hotspots, negative bowls also naturally occur when a
uniform disc with spots is fitted by a disc alone. Negative
patches towards the star can also occur owing to an increase of opacity along
the line of sight. In the limb these patches can be the result of a
convolution of the beam with the residuals that remain when a top-hat
stellar disc is subtracted from a more shallow stellar profile. In the Band 4
observations, we find that the prominent hotspot can be constrained to
$<5.3\times5.3$~mas, which implies $T_{\rm b}>3750$~K. In Band 7, the two
spots are very compact ($<2.6\times2.6$~mas) with $T_{\rm b}>10000$~K. This
is consistent with the previously observed hotspot. We find no
correspondence in the positions of the spots seen in Bands 4 and 7, nor with the position of the spot seen in previous
observations. Since the region probed at the shorter wavelengths (at
$\sim1.2~R_*$) is still optically thick at longer wavelengths (where
we find a radius of $\sim1.33~R_*$), the Band 7 spots are likely
obscured at Band 4. Additionally, the current observations were taken
almost a month apart, while the previous observations were taken two
years earlier. Since significant changes in the optical are seen on
timescales of weeks (Khouri et al., in prep.) and
convective motions and associated shocks operate on similar
timescales. it is likely that the lifetime of the hotspots is of the
same order.

{\bf R~Leo:} Although the multi-epoch observations of R~Leo show a
clear signature of an expanding layer, there is no sign of strong
hotspots.  We only see an extension in
the north-east in the Band 6 observations. This structure appears marginally resolved
($\sim10\times5$~mas implying $T_{\rm b}\sim1000$~K) along the limb of the
star, which means it could be related to the layer invoked to explain
the observed sizes and temperatures. The ellipticity also increases
from nearly circular to $\sim7\%$ ellipticity with a position angle
due north. The fact that the residuals remain towards the north-east
implies that the true shape of the stellar disc is not quite
elliptical and that an expansion of material with a size of up to an
entire quarter of the stellar disc is occurring. Such a large size would
not be inconsistent with that of a large convective cell.

\subsection{Extended emission surrounding Mira~A}
\label{excess}

\begin{figure}
\centering
\includegraphics[width=0.47\textwidth]{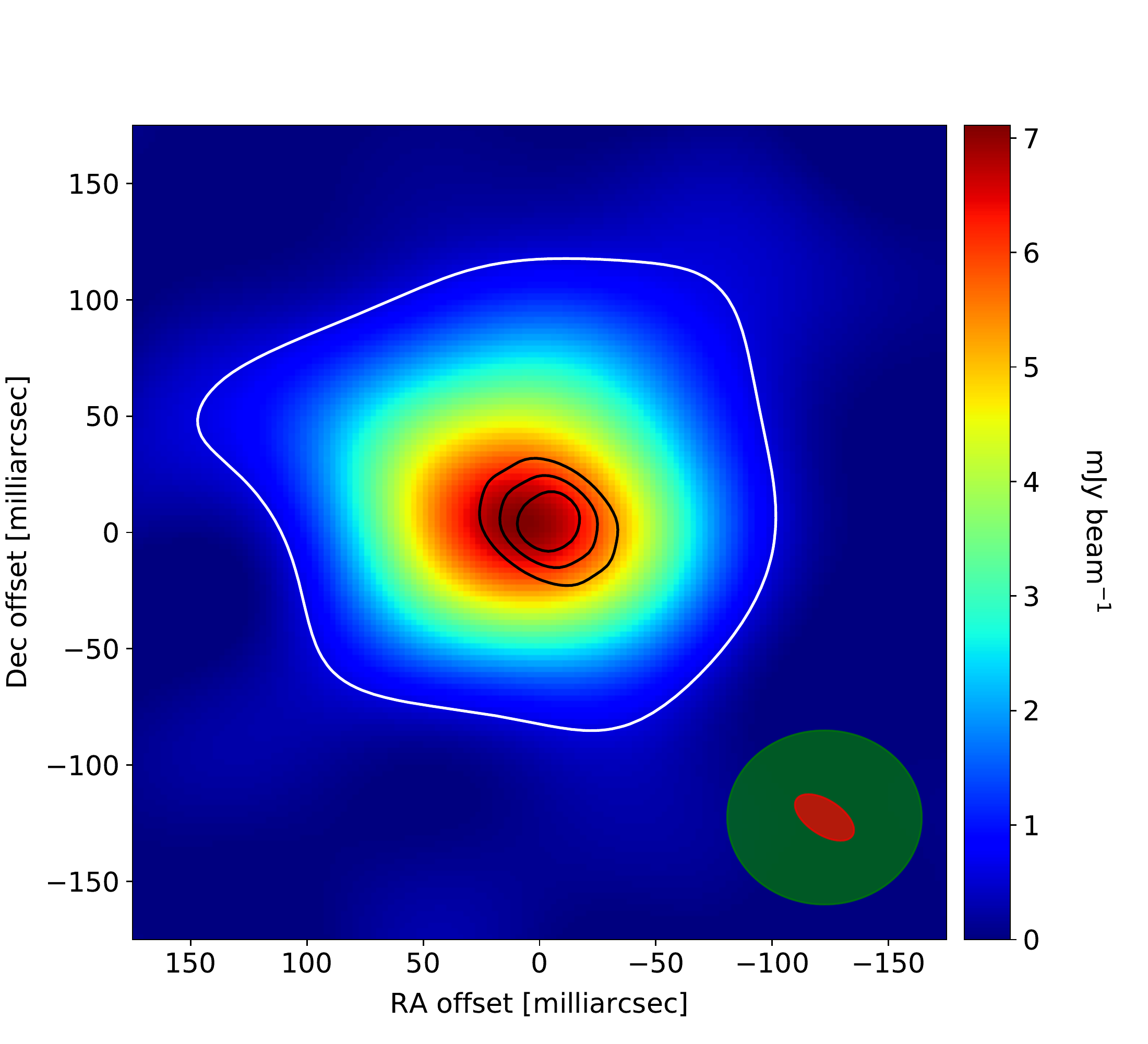}
\caption{Residual map of the extended emission around Mira A at
  332~GHz. We have subtracted the best-fit uniform disc model after fitting to the
  data with {\it uv} distance $>3$~M$\lambda$. The residual image is made
  using Briggs weighting and restricting the baselines to
  $<2$~M$\lambda$, which results in a synthesised beam of
  $84\times75$~mas at a position angle of $89.6^\circ$ (indicated in
  green in the bottom right corner). The black
  contours are the corresponding map of Mira~A using all data drawn at
  $0.2, 0.5,$ and $0.8$ times the peak intensity. The beam size for
  all data is shown in red in the bottom right corner. The white contour is
  drawn at $\sim20\sigma$, with $\sigma=0.13$~mJy~beam$^{-1}$. The
  residual emission has a peak of $7$~mJy~beam$^{-1}$, which
  corresponds to $\sim3\%$ of the stellar flux density.}
     \label{MiraExc}
\end{figure}

As noted previously, in particular the shortest wavelength Mira data
show a clear deviation from a simple stellar disc model at the largest
angular scales. We show this residual, for one of the spws, in
Fig.~\ref{MiraExc}. The same emission is seen in all spws and has an
almost linear increase of integrated flux density from $12$~mJy at $330$~GHz
to $31$~mJy at $344$~GHz. It is immediately apparent that the extended
emission peaks towards the east of the star and extends over almost
$5~R_*$. The extent and morphology bears resemblance to the emission
seen in SO, CO, and TiO$_2$ lines shown in \citet{Khouri18} and extends
out to the dust ridges seen with SPHERE/ZIMPOL in the same paper.

We suggest three possible origins of the extended emission. If the
emission originates from circumstellar dust emission close to the star
it would represent a dust mass of $\sim2\times10^{-7}$~M$_\odot$ at
$T=1000$~K adopting circumstellar dust properties from
\citet{Knapp93}. The large flux density difference between the spws would
argue against the emission being due to dust. However, since the
emission only corresponds to a few percent of the subtracted stellar
emission, the flux density measurements might be uncertain. If the
emission is due to dust, it is however unclear why the VLT/SPHERE
observations \citep{Khouri18} show such a clear lack of dust
scattering emission in the region.

A second possibility would be that the emission is due to many weak
residual spectral lines remaining in the data. Although a significant
effort was made to remove the obvious lines, the line density at the
shortest wavelengths around Mira A is significant. If we are seeing
stacked lines this could explain the correspondence with the
aforementioned stronger lines arising in the same region. It would
hint at an asymmetric mass-loss process that is not surprising
considering the complexity of the Mira system. It is not 
clear why the flux density of the emission changes almost linearly with
frequency.

Finally, the emission could represent the electron-neutral free-free
emission from the optically thin extended atmosphere. Although we find
that the brightness profile is generally well represented by a stellar
disc model, the true profile could have weaker emission extending out to
several stellar radii. The Mira 46~GHz observations indicate an extent of almost
$4~R_*$. Furthermore, although we were unable to produce a similar
high fidelity map at Band 6, at those frequencies a
short-baseline excess was also found.  This hypothesis does not
naturally explain the strong flux density increase with frequency, but
potentially rapid changes of opacity could be involved. If this is
indeed the cause for the emission, it would also indicate quite a
non-symmetric envelope.

At the moment we cannot rule out any of the possible origins with
certainty, although dust emission appears unlikely. To
  distinguish between these possibilities,
sensitive observations at high frequencies and improved models are
needed.

\section{Conclusions}
\label{conc}

We have shown that multi-epoch and multi-frequency high
angular-resolution observations with ALMA can be used to probe the
extended atmospheres of AGB stars. This has allowed us to constrain
the density, ionisation, and temperature between $\sim1-2~R_*$.  The
main source of uncertainty on the temperature determination has been
shown to be the absolute flux calibration accuracy of
ALMA. Specifically, multi-epoch observations of the same flux- and
gain-calibrator pair over four epochs of ALMA Band 4 observations
indicate that an uncertainty of $\sim5\%$ can only be reached with
sufficient observations within a short timescale. For single epoch
observations, the uncertainty, even at low frequency, is closer to
$10\%$. Still, size measurements and surface maps can be made with
great accuracy. 

We find that parameterised atmosphere models can
describe the observations for typical surface hydrogen number
densities between $n_{\rm H_2,0}=3\times10^{12} -
1\times10^{13}$~cm$^{-3}$ and photospheric temperatures of
$T=2300-3000$~K. These models assume the electron-neutral free-free
process as the dominant opacity source as previously found in
RM97. However, for R~Dor, Mira~A, and R~Leo a layer of increased
continuum opacity is needed to reproduce the shallow slope of the size
versus frequency relation. This layer, of unknown thickness, could be
due to a sudden increase of neutral hydrogen density or effective
ionisation or a combination of this. Alternatively an unknown
additional opacity source is needed. It is likely that the layer is
non-uniform and it might be related to convective shocks.

A mottled structure of spots can be seen most clearly on the surface
of R~Dor and W~Hya. The brightness temperature of the brightest spot
of R~Dor is $>3000$~K while the two brightest spots on the surface of
W~Hya have $T_{\rm b}>10000$~K, which is consistent with previous
observations \citep{Vlemmings17}. The spots likely represent the
shocks that cause the formation of a low-filling factor
chromosphere. The timescale of these spots is of the order of a few
weeks at most.

Our six epochs of observations of R~Leo show a linear increase in
measured size with an expansion velocity of
$10.6\pm1.4$~km~s$^{-1}$. We suggest that the expansion is related to
the motion of the aforementioned high opacity layer. The velocity is
consistent with the typical line widths seen in the stellar atmosphere
but does not exceed the escape velocity.

In particular for Mira~A, we find significant excess flux at the
shortest baseline lengths compared to a uniform stellar disc
model. The most likely explanation is that we are seeing either an
asymmetric distribution of optically thin neutral-ion free-free
emission out to almost $3$~R$_*$, or the effect of stacking weak
emission lines in what were thought to be line-free regions of the
spectrum. However, although unlikely, we cannot yet fully rule out
that the emission comes from warm circumstellar dust.

\begin{acknowledgements}
  This work was supported by ERC consolidator grant 614264. WV, TK,
  and HO also acknowledge support from the Swedish Research
  Council. This paper makes use of the following ALMA data:
  ADS/JAO.ALMA\#2016.1.00004.S, ADS/JAO.ALMA\#2016.A.00029.S,
  ADS/JAO.ALMA\#2017.1.00075.S, ADS/JAO.ALMA\#2017.1.00191.S and
  ADS/JAO.ALMA\#2017.1.00862.S. ALMA is a partnership of ESO
  (representing its member states), NSF (USA) and NINS (Japan),
  together with NRC (Canada), NSC and ASIAA (Taiwan), and KASI
  (Republic of Korea), in cooperation with the Republic of Chile. The
  Joint ALMA Observatory is operated by ESO, AUI/NRAO and NAOJ. We
  also acknowledge support from the Nordic ALMA Regional Centre (ARC)
  node based at Onsala Space Observatory. The Nordic ARC node is
  funded through Swedish Research Council grant No 2017-00648.
\end{acknowledgements}

%
%


\end{document}